\documentclass[12pt,manuscript]{aastex61}

\usepackage{amsmath,amstext}
\usepackage{apjfonts} 
\usepackage{multirow}

\newcommand{\kms}{\ensuremath{{\rm km~s}^{-1}}}

\newcommand{\ha}{H$\alpha$}

\newcommand{\hh}{H$_{2}$}
\newcommand{\hi}{\ion{H}{1}}
\newcommand{\hii}{\ion{H}{2}}

\newcommand{\feii}{[\ion{Fe}{2}]}

\shorttitle{Kinematic Distances to Galactic Supernova Remnants}
\shortauthors{Lee et al.}

\begin{document}

\title{High-resolution Near-infrared Spectroscopic Study of
Galactic Supernova Remnants. I. Kinematic Distances}

\author[0000-0003-3277-2147]{Yong-Hyun Lee}
\email{yhlee.astro@gmail.com}
\affiliation{Korea Astronomy and Space Science Institute,
Daejeon 305-348, Korea}
\affiliation{Department of Physics and Astronomy, Seoul National University,
Seoul 151-747, Korea}
\author[0000-0002-2755-1879]{Bon-Chul Koo}
\affiliation{Department of Physics and Astronomy, Seoul National University,
Seoul 151-747, Korea}
\author[0000-0003-0894-7824]{Jae-Joon Lee}
\affiliation{Korea Astronomy and Space Science Institute,
Daejeon 305-348, Korea}

\begin{abstract}
We have carried out
high-resolution near-infrared spectroscopic observations
toward 16 Galactic supernova remnants (SNRs)
showing strong \hh\ emission features.
A dozen bright \hh\ emission lines are clearly detected
for individual SNRs,
and we have measured their central velocities, line widths, and fluxes.
For all SNRs except one (G9.9$-$0.8),
the \hh\ line ratios are well consistent with that of thermal excitation
at $T\sim2000$~K,
indicating that the \hh\ emission lines are
most likely from shock-excited gas 
and therefore that they are physically associated with the remnants.
The kinematic distances to the 15 SNRs are derived from
the central velocities of the \hh\ lines
using a Galactic rotation model.
We derive for the first time the kinematic distances to four SNRs: 
G13.5$+$0.2, G16.0$-$0.5, G32.1$-$0.9, and G33.2$-$0.6.
Among the remaining 11 SNRs, 
the central velocities of the \hh\ emission lines for six 
SNRs are well consistent ($\pm5~\kms$) with 
those obtained in previous radio observations, while for the other 
five SNRs (G18.1$-$0.1, G18.9$-$1.1, Kes 69, 3C 396, W49B)
they are significantly different. 
We discuss the velocity discrepancies in these five SNRs. 
In G9.9$-$0.8, the \hh\ emission shows nonthermal line ratios
and narrow line width ($\sim 4~\kms$),
and we discuss its origin.
\end{abstract}


\section{Introduction} \label{sec-int}
Distance is an essential and important parameter for the study of
Galactic supernova remnants (SNRs),
but its determination is difficult as in most areas of astronomy.
Among the $\sim 300$ known Galactic SNRs, therefore,
only a limited number of SNRs have reliable distance estimates
\citep{green19}.

A most popular distance determination method for SNRs is
to derive the kinematic distance from velocity information
using a Galactic rotation model. 
The classical method is to observe \hi\ 21~cm absorption,
which was introduced in the mid-1950s   
and was first applied to the SNR Cassiopeia A 
(\citealt{wil54,hag55}; see also \citealt{mul59}).
In this \hi\ absorption technique, 
the absorption of 21 cm continuum from an SNR
by intervening {\it cold} \hi\ clouds is observed, and 
the maximum velocity where the absorption occurs provides a 
{\it lower and/or upper} limit to the distance to the SNR.
For SNRs bright in radio,
it is straightforward to apply this technique
because the absorption profiles can be obtained directly from
either single-dish or interferometric observations, 
although the noncircular motions of \hi\ gas 
due to streaming along the spiral arms 
can hamper the derivation of accurate limits
\citep[e.g.,][]{cla62,green89,koo93,kothes03,kothes13}.
For SNRs faint in radio, however, it is not straightforward because,
in order to obtain the absorption spectrum by an SNR,
the background spectrum needs to be derived from
the observation of the surrounding area and subtracted from the SNR spectrum,
where the random fluctuation of \hi\ profiles along different lines of sight
could produce artificial absorption features \citep[e.g.,][]{green82}. 
Recently, an improved method using high-resolution \hi\ data,
together with CO observation data, has been devised
and systematically applied to two dozen SNRs
\citep[][and references therein]{lea10,ran18b,ran18a}.

For SNRs interacting with ambient molecular clouds (MCs),
their kinematic distances can also be derived from the velocities of the MCs.
According to the compilation by \citet{jia10} and \citet{kil16},
there are $\sim 50$ SNRs suggested to be interacting with MCs.
The evidence for the interaction ranges from a simple morphological association
to the presence or detection of shock signatures such as
the broad molecular lines, the high $^{12}$CO $J=$2--1/1--0 line ratios, 
1720 MHz OH masers, and near-infrared (NIR) rovibrational \hh\ emission lines.
The kinematic distances to SNRs have been derived mainly from
the velocities of CO emission lines
\citep[][and references therein]{jia10,kothes03,kil16}
or OH masers \citep[][and references therein]{fra96,gre97,yus03,hew08}.
The studies using \hh\ emission lines are relatively sparse.
The \hh\ 2.122~\micron\ line had been first detected in the SNR IC~443 in 1979
\citep{treffers79}, and later in several well-studied SNRs
(see \citealt{lee19} and references therein).
There have, however, been neither systematic studies of SNRs
in the \hh\ 2.122~\micron\ line nor spectroscopic studies 
to measure the distance of SNRs using the \hh\ 2.122~\micron\ line
until very recently (see below).
This contrasts with CO $J=$1--0 115 GHz or OH 1720 MHz maser lines,
the importance of which as a distance indicator had been realized early
and for which systematic surveys toward SNRs were carried out
\citep[e.g.,][]{huang86,fra96,gre97}.
The major reason for the sparseness of such studies
in the \hh\ 2.122~\micron\ line
might be because high-performance NIR detector arrays
became available relatively recently \citep[e.g.,][]{rieke07}.

Recently, we have carried out a systematic \hh-emission line study of 
SNRs in the inner Galaxy using 
the UKIRT Wide-field Infrared survey for \hh\ \citep[UWISH2;][]{fro11},
which is an unbiased survey of the inner Galactic plane
($7\degr < l < 65\degr$ and $|b| < 1.3\degr$)
in the \hh\ 2.122~\micron\ emission line \citep{lee19}.
Among the 79 Galactic SNRs in the survey area,
we detected a total of 19 SNRs with \hh\ emission features.
Comparing the \hh\ and radio continuum morphologies,
we suggested that the \hh\ emission features are associated with the SNRs.
Some \hh\ emission features are found outside
the \feii/X-ray/radio boundaries
(known as the ``\feii-\hh\ reversal'' phenomena),
but their spatial distribution, together with
the \hh\ emission-line ratios from
a follow-up NIR spectroscopy, indicates that
these \hh\ features are also probably associated with the SNRs
(see also Section~\ref{sec-res-ext}).

In this paper,
we present results from high-resolution NIR spectroscopy of
the \hh\ emission features around the Galactic SNRs detected by \citet{lee19}.
The purpose of the study is to derive their kinematic distances
by measuring accurate central velocities of \hh\ lines.
This paper is organized as follows.
In Section~\ref{sec-obs},
we describe our NIR spectroscopic observations and the data reduction procedure.
In Section~\ref{sec-res},
we examine the physical association of 
\hh\ emission with the SNRs by  
investigating the excitation mechanism of the \hh\ emission
from their line ratios and line widths.
We also derive their kinematic distances from the central velocities
and compare them with those of previous radio observations.
In Section~\ref{sec-dis}, we discuss 
one SNR showing nonthermal excitation of \hh\ emission lines
and also five SNRs whose \hh\ central velocities are
significantly different from previous results.
The summary of this paper is given in Section~\ref{sec-sum}.

\section{Observations and Data Reduction} \label{sec-obs}
\subsection{Near-infrared Spectroscopy} \label{sec-obs-nir}
We carried out NIR spectroscopy of \hh\ emission-line features
detected around the Galactic SNRs
using the Immersion GRating INfrared Spectrograph (IGRINS)
mounted on the Harlan J. Smith 2.7~m telescope located at McDonald Observatory.
IGRINS is a high-resolution (resolving power $R\sim45{,}000$)
NIR spectrograph
that provides simultaneous wavelength coverage of {\it H} and {\it K} band
\citep[1.5--1.8~\micron\ and 1.9--2.5~\micron, respectively;][]{yuk10,par14}.
The slit width and length are $1\arcsec$ and $15\arcsec$, respectively.

The observations had been carried out in 2014 May, 2015 June, and 2016 July.
Among the 19 \hh-emitting SNRs confirmed by \citet{lee19},
16 were observed (Figure~\ref{fig-slitpos-1}).
We carefully selected slit positions to obtain spectra of
strong \hh\ emission features at the border of the remnants.
For several SNRs, the spectra at multiple slit positions had been obtained. 
The slit positions of all the targets are shown in Figure~\ref{fig-slitpos-1}.
The sizes of the emission features are either comparable to
or larger than the slit length,
so we took ``OFF''-position spectra
just before or after the target observations
to subtract the sky-background emission/continuum.
The single exposure time per frame was limited to 300 s
in order to prevent saturation of the bright-sky airglow emission lines.
To increase the sensitivity for weak \hh\ emission-line features,
however, we took multiple observations for several \hh\ emission features.
We also took several spectra of nearby A0V standard stars
for photometric calibration.

For the data reduction,
we utilized the dedicated pipeline for IGRINS written in
{\sc Python}\footnote{PLP version 2.2.0. The pipeline package is downloadable at
\href{https://github.com/igrins/plp}{https://github.com/igrins/plp}}.
The pipeline performs flat-fielding correction and sky subtraction
at the beginning of the data reduction.
Then,
it performs distortion correction to produce two-dimensional spectra.
Wavelength calibration was done by
comparing the observed OH airglow emission lines with their vacuum wavelengths,
and the overall $1 \sigma$ uncertainty is about half-pixel width,
which corresponds to $\sim 1~\kms$.
For photometric calibration together with telluric absorption correction,
we derived the wavelength-dependent conversion factor
from digital to physical units
by comparing the observed A0V standard stars with the Kurucz model
spectra\footnote{Kurucz, R. L. 2003,
\href{http://kurucz.harvard.edu/}{http://kurucz.harvard.edu/}},
and we multiplied it by each target spectrum.

In all the spectra,
we clearly detected bright \hh\ 2.122~\micron\ emission lines.
Most of them show other \hh\ emission lines in $H$ and $K$ bands too.
We performed a single Gaussian fit for all the detected \hh\ lines
(Figure~\ref{fig-spec}),
except for the lines detected in four slits
(G18.9$-$1.1-C, Kes 69-NE, and 3C 391-NE and SW),
to derive their central velocities, line widths, and fluxes.
For the four exceptions,
the observed lines are broad
and not well fitted by a single Gaussian shape
(Figure~\ref{fig-spec}).
In such cases, we first 
performed fitting the \hh\ 2.122~\micron\ line
with two Gaussian components,
and then we applied its profile to the remaining \hh\ lines
to measure their fluxes.
Therefore, the relative flux ratio of the two velocity components is the same
for every \hh\ line.
All the observed central velocities have been corrected to the velocities
with respect to the local standard of rest (LSR; $v_{\rm LSR}$),
and the observed line widths have been corrected for the 
instrumental broadening (7~\kms).
The extinction correction for the observed line fluxes has been done by
using the hydrogen column density ($N_{\rm H}$)
derived from previous studies (see Table~4 of \citet{lee19}),
adopting a general interstellar extinction model with $R_{V}=3.1$
\citep{dra03}.
For four SNRs, G9.9$-$0.8, G13.5$+$0.2, G16.0$-$0.5, and G33.2$-$0.6,
however, $N_{\rm H}$ are not available,
so we estimated the visual extinction $A_{V}$ from
the kinematic distance derived in this work
by using the average ratio of visual extinction to path length
in the solar neighborhood
$\left\langle A_{V}/L \right\rangle \approx 1.8$ mag kpc$^{-1}$
\citep[][]{whittet92}
and converted it to the NIR extinction using the extinction model of
\citet{dra03}.
The derived physical properties are listed in Table~\ref{tab-prop}.

\subsection{$^{12}CO$ $J=1$--$0$ Archival Data} \label{sec-obs-co}
In order to search MCs associated with the SNRs,
we utilize the high-resolution $^{12}$CO $J=1$--$0$ data
from the FOREST (FOur-beam REceiver System on the 45 m Telescope)
Unbiased Galactic Plane Survey
with Nobeyama 45 m telescope \citep[FUGIN;][]{ume17},
which is one of the legacy projects
utilizing the new multibeam receiver FOREST.
The receiver has a wide bandwidth to obtain $J=1$--$0$ emission-line data of
$^{12}$CO, $^{13}$CO, and C$^{18}$O simultaneously.
The survey was carried out from April 2014 to March 2017
to cover the first ($10\degr < l < 50\degr$, $|b| < 1\degr$)
and the third ($198\degr < l < 236\degr$; $|b| < 1\degr$)
quadrants of the Galactic plane.
The telescope has a beam size of $\sim 15\arcsec$ at 115 GHz,
and the mapping was done with a sampling interval of $8\farcs5$.
The archival data provide an effective angular resolution of $20\arcsec$
and velocity resolution of 1.3~\kms.
The typical sensitivity of the $^{12}$CO data is 0.24~K.
The calibrated data cubes in FITS format were retrieved from
the Japanese Virtual Observatory portal
\footnote{\url{http://jvo.nao.ac.jp/portal/}} operated by
the Astronomy Data Center of the National Astronomical Observatory of Japan.

\section{Results} \label{sec-res}
\subsection{Association with SNRs} \label{sec-res-ext}
The \hh\ emission features in the 16 SNRs in Table~\ref{tab-prop} 
are ``morphologically'' associated with the SNRs \citep{lee19}.
We, however, cannot rule out the chance projection of the
\hh\ emission features on the SNRs.
In order to confirm their physical association,
we investigate the excitation mechanism of the \hh\ emission 
using \hh\ line ratios.
Previous NIR spectroscopic studies of several SNRs have shown that 
the \hh\ lines excited by SNR shocks have 
thermal line ratios corresponding to $T \sim 2000$--$3000$~K
(IC 443, \citealt{burton88,burton89,moorhouse91};
RCW 103, \citealt{oliva90,burton93};
Cygnus Loop, \citealt{graham91};
G11.2$-$0.3, \citealt{koo07,lee19};
Kes 69, \citealt{lee19}).
In this paper, we use the flux ratios of
2--1 S(1)/1--0 S(1), 1--0 S(2)/1--0 S(0), and 1--0 S(1)/1--0 S(0)
(Table~\ref{tab-prop})
to confirm the physical association
between the observed \hh\ lines and the SNRs.

The flux ratio of 1--0 S(1) 2.122~\micron\ and 2--1 S(1) 2.248~\micron\ lines
has been widely used to distinguish thermal and nonthermal excitation.
\citep[e.g.,][]{bur92,mou94,pak04,maz13,le17}.
Since these two lines originate from the same rotational state
but from different vibrational states,
their flux ratio is independent of the ortho-to-para ratio (OPR) 
that could affect the observed line ratio.
For the collisionally excited emission lines from
warm ($T=2000$--3000 K) \hh\ gas,
the flux ratio is in the range of 0.1--0.2 \citep[e.g.,][]{mou94}.
Nonthermal excitation by UV fluorescence, on the other hand, 
involves the UV pumping of \hh\ molecules to the excited electronic 
states and the cascade to the lower states, and the \hh\ line ratios
depend on the branching ratios in the downward cascade.
This increases the populations of the high-energy levels ($v\geq2$),
which yields a flux ratio 2--1 S(1)/1--0 S(1)
higher than that from the collisionally excited \hh\ gas.
For the pure UV fluorescence emission from  
the diffuse \hh\ gas ($\lesssim 10^{4}$ cm$^{-3}$)
irradiated by UV photons,
the 2--1 S(1)/1--0 S(1) ratio is in the range of 0.5--0.6
\citep{bla87,ste89,bur92}.
The ratios among the $v=1$--0 rovibrational lines in     
UV fluorescence emission 
are also different from those in 
thermal emission, so that the flux ratios 
1--0 S(2)/1--0 S(0) and 
1--0 S(1)/1--0 S(0) are also useful 
to discriminate between thermal and nonthermal excitation. 
At higher densities, a substantial column of UV-heated, warm 
gas develops and the thermal emission dominates the $v=1$--0 lines,
but the temperature of this UV-heated thermal gas is
$T\le 1000$~K \citep{ste89}, substantially lower than 
that of the shocked gas, 
so that the UV-heated thermal gas is clearly separated from  
the shock-excited gas in the \hh\ line ratio diagrams
\citep[e.g.,][]{mou94}.

Figure~\ref{fig-ratio} shows the distribution of \hh\ line ratios of 34   
positions in 16 SNRs 
where the abscissa is 2--1 S(1)/1--0 S(1) 
and the ordinate is 1--0 S(2)/1--0 S(0) or 1--0 S(0)/1--0 S(1). 
The solid line represents the locus of 
thermal emission from a source in local thermodynamic equilibrium (LTE). 
In the figure, we also mark the locations of nonthermal UV fluorescence 
emission \citep[][]{bla87} and UV-heated thermal emission \citep{ste89}.
Figure \ref{fig-ratio} shows that most data points 
are located around the $T\sim 2000$~K in LTE,
consistent with collisionally excited thermal emission from shocked gas.
Their line ratios are significantly different from 
either the UV fluorescence emission or the UV-heated thermal emission.
There are several positions with upper limits in 2--1 S(1)/1--0 S(1)
(i.e., G11.2$-$0.3-N, G13.5$+$0.2-S, G16.0$-$0.5-E1 and E2, G18.1$-$0.1-SE, 
G32.1$-$0.9-NW and C, G33.2$-$0.6-W, and 3C 396-W2),
but their upper limits are much smaller than
that of the pure nonthermal UV fluorescence
\citep[0.50--0.60;][]{bla87}, and 
their 1--0 S(2)/1--0 S(0) and/or 1--0 S(0)/1--0 S(1)
are significantly different from that of the UV-heated thermal emission. 
Therefore, the 1--0 S(1) \hh\ emission lines from those positions are   
also probably collisionally excited thermal emission (see below).
The only exception is G9.9$-$0.8-NW.
Its 2--1 S(1)/1--0 S(1) ratio is $0.31\pm 0.04$,
which is clearly larger than that of the thermal emission at 2000 K.
But the ratio is also considerably smaller than the typical line ratio of 
the pure UV fluorescence emission.
Its 1--0 S(0)/1--0 S(1) ratio ($0.49\pm0.04$) is also much larger than 
that of the thermal emission
and is almost comparable to those of the pure nonthermal UV fluorescence
\citep[0.45--0.60;][]{bla87}.
These line ratios suggest that the \hh\ emission in G9.9$-$0.8-NW 
is likely a mixture of thermal and nonthermal emission.
It is worthwhile to note that its line width 
is also very small ($4.4\pm 0.1$~\kms) compared to the other 
SNRs, whose line widths are larger than $9~\kms$.
It is comparable to the line width \citep[2--6~\kms;][]{kap17,le17,oh18}
of nearby photodissociation regions (PDRs).
We will discuss the nature of this source in Section~\ref{sec-dis-g9.9}.

One thing to note in Figure~\ref{fig-ratio} is that there is a large scatter 
in the line ratios around the solid line
representing the thermal \hh\ gas in LTE conditions.
This might be mostly due to the uncertainty in the 
1--0 S(0) and 1--0 S(2) line fluxes, although it could also be partly  
due to the uncertainty in extinction correction.
The weighted means for the line ratios are
2--1 S(1)/1--0 S(1) = $0.101\pm 0.003$, 
1--0 S(2)/1--0 S(0) = $1.98\pm 0.03$, and 
1--0 S(0)/1--0 S(1) = $0.188\pm 0.003$, which 
agree well with the thermal emission at 2000~K. 
It is, however, worthwhile to note that
several \hh\ emission features are beyond the SNR boundary in radio:
G11.2$-$0.3-N, S, SE, and NE; Kes 73-W; W44-N; 3C 396-W1 and W2;
and W49B-E1 and E2 (Figure~\ref{fig-slitpos-1}; see also \citealt{lee19}).
Most of them have some morphological connection with the SNRs,
i.e., they are either connected to an SNR filament (W44) or 
parallel to the SNR boundary (Kes 73, 3C 396, and W49B).
This, together with their thermal line ratios,
strongly supports their physical association with the SNRs.
Their central velocities are also comparable to
those of the other positions in individual SNRs (Table~\ref{tab-prop}).
Perhaps the only exception is G11.2$-$0.3.
In this SNR, the \hh\ emission features extend far beyond the SNR boundary
and there is no obvious morphological connection
between the \hh\ emission features and the SNR.
We can find no \hii\ region spatially coincident with the \hh\ emission
in the WISE Catalog of Galactic \hii\ regions
\citep[V2.2;][]{anderson14}.
On the other hand, the line ratios of all four positions are consistent with
the shock excitation, and their central velocities are comparable to
the suggested systemic velocity of the SNR \citep[$+45$~\kms;][]{gre88},
so that the association with the SNR is likely (see also \citealt{koo07}).
Such \hh\ filaments have been known from the 
early days of NIR observations, and 
it was pointed out  that heating/exciting sources other than 
the SNR shock are required for the \hh\ emission
\citep{oliva90,graham91,burton93}.
One possibility is the X-ray emission from SNRs;
if molecular gas is irradiated by X-ray, the temperature can reach $\le 3000$ K 
and the \hh\ line ratios can be close to those of shocked gas
\citep{lep83,dra90}.
We will explore this possibility and 
the origin of such \hh\ emission in our forthcoming paper.  
In this work, we will not distinguish those positions from the rest,
which is acceptable because their velocities are 
comparable to those of the other positions in an SNR.

\subsection{Kinematic Distances} \label{sec-res-dis}
Our analysis in Section~\ref{sec-res-ext} suggests that
the observed \hh\ emission features are all physically associated with
the SNRs except one (G9.9$-$0.8). 
The central velocities of the \hh\ 2.122~\micron\ emission lines
in Table~\ref{tab-prop}, therefore, 
should be close to the systemic velocities of the SNRs.
For the \hh\ emission features in the central areas of the SNRs, however,
there could be a large contribution from the shock motion
to the central velocity.
Indeed, in G32.1$-$0.9 and W44, 
the central velocities from the central area slits are
either much ($\sim 20~\kms$) larger (G32.1$-$0.9) or smaller (W44)
than those from the other slits.
We therefore exclude the data from the central area slits 
(i.e., slit name with ``C'' in Table~\ref{tab-prop})
in deriving the systemic velocities.
We also exclude the data from the slits
showing two velocity components (see Figure~\ref{fig-spec}), 
although one of the two components could represent the 
systemic velocity of the SNR. 
We then proceed to obtain the systemic velocities of the SNRs as follows:
(1) For the SNRs with a single slit observation,
we simply take the central velocities.
(2) For the SNRs with multiple slit observations and 
with the spread in velocity less than 5~\kms, which is comparable to
the cloud-to-cloud velocity dispersion of MCs 
in the inner Galaxy \citep[e.g., see Figure~4 of][]{clemens85},
we take the average of the central velocities
weighted by 1/$\sigma_{v}^{2}$,
where $\sigma_{v}$ is the $1\sigma$ uncertainty of the central velocity
from the Gaussian fitting (Section~\ref{sec-obs-nir}).
(3)  For the SNRs with multiple slit observations and 
with velocity spread larger than 5~\kms, 
we present the range given by the minimum and maximum 
central velocities.

Table~\ref{tab-dis} shows the systemic velocities ($v_{\rm LSR}$) of
the 16 SNRs derived from our \hh\ emission lines. 
Note that we have included SNR G9.9$-$0.8 in parentheses, although 
the association of the H$_2$ emission with the SNR is not certain. 
The table also shows the systemic velocities determined from 
\hi\ absorption, CO emission, or OH maser observations in previous studies.
For four SNRs (G13.5$+$0.2, G16.0$-$0.5, G32.1$-$0.9, and G33.2$-$0.6),
no velocity information can be found in the literature. 
The left panel of Figure~\ref{fig-vel} compares the systemic velocities
of 12 SNRs from our work with
those from previous studies in radio ($v_{\rm LSR, radio}$).
For seven SNRs, they agree with each other within $\pm5~\kms$, 
supporting the association of all emission features with the SNRs.
On the other hand, for five SNRs  
(G18.1$-$0.1, G18.9$-$1.1, Kes 69, 3C 396, and W49B)
the discrepancy between the two velocities is large 
(12--52~\kms).
There could be several possible explanations.
First, the \hh\ emission-line velocities could have
some contribution from shock motion,
although it might be small because the \hh\ filaments are located
near the edge of the SNRs. 
Alternatively, the velocities obtained in previous studies,
mostly from \hi\ and/or CO observations,
can be in substantial error.
In Section~\ref{sec-dis},
we will discuss the origin of the discrepancies in these five SNRs.

The $v_{\rm LSR}$ derived either 
from our \hh\ emission lines or from radio observations
can be converted to kinematic distance,
adopting an appropriate Galactic rotation curve model.
We use the recent model of \cite{rei14}, which is based on 
$\sim 100$ high-mass star forming regions 
with accurate distances determined by trigonometric parallax.
The model adopts the `universal' rotation curve of \citet{per96}
with new Galactic parameters ($R_{0}=8.34$ kpc and $V_{0}=241~\kms$).
The kinematic distances obtained by using the 
model of \citet{rei14} are listed in Table~\ref{tab-dis}.
For the SNRs
without previous systemic velocity measurements,
we present the distances estimated from other nonkinematic 
methods such as the $\Sigma$-$D$ relation,
where $\Sigma$ and $D$ are the radio surface brightness and
the diameter of the SNRs
or the Sedov analysis for comparison (see Table~\ref{tab-dis}).
The Sedov analysis assumes that the SNR is in the Sedov phase and
has a canonical SN explosion energy of $10^{51}$~erg.
These distance estimates can be used but have a large uncertainty.
For G18.9$-$1.1, we also give the distance (1.8 kpc) derived from 
the comparison of the X-ray absorbing columns 
and the extinction variation to the SNR direction 
\citep{shan18}.
It is worth noticing that there are two kinematic 
distances, near-side and far-side distances, 
corresponding to a $v_{\rm LSR}$.  
For the majority of SNRs, this distance ambiguity has been resolved
by \hi\ absorption studies (see Section~\ref{sec-int}).
For the SNRs without such information, 
we have adopted the kinematic distances close to those from 
the $\Sigma-D$ relation or the Sedov analysis.   
They are marked by question marks in 
Column (3) of Table~\ref{tab-dis}. 
%
The right panel of Figure~\ref{fig-vel} 
compares the  
distances derived from our NIR \hh\ observations 
with those from previous studies. 
Not surprisingly,
the five SNRs with a large discrepancy in $v_{\rm LSR}$ 
(G18.1$-$0.1, G18.9$-$1.1, Kes 69, 3C 396, and W49B)
also show large distance discrepancy ($\Delta d > 1$~kpc).
For the four SNRs 
(G13.5$+$0.2, G16.0$-$0.5, G32.1$-$0.9, G33.2$-$0.6),
previous distance estimates have large uncertainties, and our result provides 
reliable kinematic distances for the first time,
although the near-far distance ambiguity needs to be 
resolved for them.

\section{Discussion} \label{sec-dis}
In this section,
we discuss the origin of the dense molecular gas
emitting \hh\ emission lines detected in the northwestern border of G9.9-0.8.
We also discuss the five SNRs
(G18.1$-$0.1, G18.9$-$1.1, Kes 69, 3C 396, and W49B)
showing a large discrepancy in $v_{\rm LSR}$
between our NIR \hh\ emission-line observations
and the previous radio observations.

\subsection{G9.9$-$0.8} \label{sec-dis-g9.9}
SNR G9.9$-$0.8 was first identified by \citet{bro06}
in their VLA 90~cm observation
and shows a broken shell-like morphology with a radius of 6\arcmin.
\citet{stu11} detected strong \ha\ emission features
in the northwestern border of the remnant
and suggested that the emission is associated with the remnant
based on the positional coincidence of
the bright radio continuum and the \ha\ emission.
\citet{kil16}
detected broadened CO emission at $v_{\rm LSR}=+27$--$+33~\kms$
near the southwestern border of the remnant and obtained
the kinematic distance of $\sim 4$~kpc
assuming that the broadening is due to the interaction with the SNR.

In \citet{lee19},
we found narrow \hh\ filaments in the northwestern border of the remnant
where the radio continuum is bright and optical \ha\ emission is detected
\citep{bro06,stu11}.
In our NIR spectroscopy of the filaments,
we found very narrow \hh\ emission lines
(FWHM $\sim 4~\kms$; Table~\ref{tab-prop})
with nonthermal line ratios (Figure~\ref{fig-ratio}),
indicating that they are likely arising from unshocked interstellar medium (ISM)
irradiated by nearby UV sources (Section~\ref{sec-res-ext}).
We also check the level population diagram of the \hh\ emission
as shown in Figure~\ref{fig-g99_lev}.
In the figure, we clearly see that
the populations are not explained by thermal excitation models
at temperatures of 2000--4000 K.
Instead, they show ``zigzag'' patterns between ortho- and para-levels.
We estimate the OPR of the \hh\ gas
using two \hh\ emission lines, 1--0 S(1) and 1--0 S(0).
Since the extinction-corrected 1--0 S(1)/1--0 S(0) is $2.1\pm0.2$,
assuming the rotational temperature of $\sim 1700$ K
from 1--0 S(2)/1--0 S(0) of $1.6\pm0.2$,
the OPR of the \hh\ gas is $\sim 1.4$,
which is much smaller than 3.0
expected in thermal \hh\ gas in LTE condition.
The zigzag patterns in the level population diagram,
together with low OPR, seem to be consistent with
typical characteristics of nonthermal UV fluorescence excitation
\citep[e.g.,][]{has87,ste99}.
But, as we pointed out in Section~\ref{sec-res-ext}, 
the observed 2--1 S(1)/1--0 S(1) ratio ($\sim 0.3$) is 
considerably smaller than that (0.6--0.7) in pure UV fluorescence excitation, 
suggesting that there could be some contribution from collisional excitation. 
The model calculations of \cite{dra96} show that
warm PDRs in low-density media can yield \hh\ line ratios
similar to what we observe in G9.9$-$0.8.
The solid line in Figure~\ref{fig-g99_lev} 
shows the line ratios obtained in one   
such model (the ``bh3d'' model of \citealt{dra96}).

In Figure~\ref{fig-g99_image},
we compare the spatial distribution of the \hh\ emission features
with that of the optical \ha\ emission features detected in \citet{stu11}.
As seen in the figure,
the \hh\ filaments are detected just outside the \ha\ emission features,
and their morphologies and distributions are almost anticorrelated.
The spatial distributions of the atomic and molecular emission lines 
suggest a PDR where 
UV photons from massive stars heat \hh\ molecules in MCs 
in contact with the \hii\ region 
\citep[][]{hol97,hol99}.
Indeed, there is an \hii\ region (G9.982$-$0.752)
around the \hh\ filaments (Figure~\ref{fig-g99_image}).
This \hii\ region was identified by \citet{loc89},
who performed a radio recombination line survey of continuum sources
in the northern Galactic plane.
Its location is well consistent with
those of the \hh\ and \ha\ emission features,
and its $v_{\rm LSR}$ (=$+34.4\pm1.5~\kms$) 
is also similar to that of the \hh\ emission line.

All the information described above indicates that
the optical \ha\ and NIR \hh\ emission is arising from the PDR
of the \hii\ region, rather than the SNR G9.9$-$0.8.
It is, however, worth noticing that
the velocity center of the broadened CO emission
\citep[+31 \kms;][]{kil16}
detected around the southwestern border of the remnant
is very close to this velocity.
If the broadened CO emission is due to
the interaction of the SNR and nearby MCs as suggested by \citet{kil16},
it is likely that
the \hii\ region and the SNR are located at nearly the same distance
($\sim 3.8$ kpc; see Table~\ref{tab-dis})
and that they are possibly associated.
Additional observations with multi-wave-band instruments may reveal
their physical connection.

\subsection{G18.1$-$0.1} \label{sec-dis-g18.1}
G18.1$-$0.1 is an SNR that belongs to the \hii\ region/SNR complex at
$l=18\fdg2$ and $b=-0\fdg3$. The complex is composed of
at least four \hii\ regions and one SNR.
G18.1$-$0.1 is located at the western edge of the complex,
and it was first confirmed as an SNR by \citet{bro06}.
In radio continuum,
it has a broken ring-like morphology
with the bright northern and eastern arcs.
Although the remnant belongs to the \hii\ region/SNR complex,
their physical association is not clear \citep[e.g.,][]{paron13,lea14}.
Using $^{13}$CO line data, \citet{paron13} suggested that
a big molecular shell at $v_{\rm LSR}=+51$--$+55~\kms$
is physically associated with the \hii\ region/SNR complex,
placing the complex at the kinematic distance of $\sim4$~kpc.
For the SNR, however,
\citet{lea14} proposed a distance of $\sim 6$~kpc based on
its \hi\ spectrum showing absorption up to $+100~\kms$,
and argued that the remnant is located at 2--3~kpc
behind of the \hii\ regions without any physical association with them.

In \citet{lee19}, we detected
strong, narrow \hh\ filaments 
at the northern and eastern borders of the remnant,
where the radio continuum brightness is enhanced \citep{bro06}.
We also found additional diffuse \hh\ emission  
outside of the remnant's western boundary,
but those emission features are likely associated with the nearby \hii\ regions.
We took NIR spectra of the northern and eastern borders of the remnant
(Figure~\ref{fig-slitpos-1}).
The morphological relation with the SNR and the line ratios
(Section~\ref{sec-res-ext}) indicate that
the \hh\ emission features are associated with the remnant.
The $v_{\rm LSR}$ of the northern filament is $+73~\kms$,
whereas that of the eastern filament is $+85~\kms$ (Table~\ref{tab-prop}).
These velocities are considerably different from 
those of previous CO/\hi\ results 
\citep[$+53~\kms$ or $+100~\kms$;][]{paron13,lea14,ran18a}.

In order to find MCs associated with the SNR,
we have investigated high-resolution $^{12}$CO $J=1$--0 data
obtained as a part of the FUGIN CO survey
(\citealt{ume17}; see also Section~\ref{sec-obs-co}).
Figure~\ref{fig-g181} displays the $^{12}$CO $J=1$--0 line channel maps
between $+48$ and $+85~\kms$.
In the channel maps,
the most prominent feature is the large MCs
between $v_{\rm LSR}=+48$ and $+56~\kms$.
In the middle of the velocity range, a ring-like structure emerges,
which is clearly seen in the $+51~\kms$ channel map.
The ring structure appears to be encircling the remnant,
with the eastern border of the remnant in contact with it.
\citet{paron13} suggested that
the \hii\ region/SNR complex is located at this velocity
and that they are physically associated with the MCs.
But there is a large gap between the ring structure and 
the northwestern shell of the SNR,
where the NIR \hh\ emission lines are detected.
The large molecular ring disappears at velocities higher than $+53~\kms$,
but instead, an additional MC filament emerges
in the northeastern outer region of the remnant.
The filament also seems to be in contact with
the eastern and northeastern boundary of the remnant,
but again its overall morphology does not match well the remnant.
From $+61$ to $+72~\kms$, on the other hand,
we see a large molecular ``wall'' that appears to be in contact with the 
eastern boundary of the remnant.
The morphology is suggestive of the interaction, but 
the velocity of the \hh\ emission in this region (+85~\kms) is 
considerably different from the CO emission.
%

At the velocities of the \hh\ emission (+72 and +85~\kms), 
no dense MCs are visible.
Instead, there are diffuse CO emission features around the remnant,
some of which are spatially coincident with the remnant (Figure~\ref{fig-g181}).
In Figure~\ref{fig-g181_zoom}, we show a close-up view of
the diffuse CO emission and compare its distribution 
with the radio continuum and \hh\ emission distributions.
The diffuse CO emission is bright along the southern boundary of the SNR,
where the radio continuum emission is faint.
There are CO `clumps' around the northwestern boundary  
of the SNR, but there is no clear spatial correlation
between these CO clumps and the northwestern \hh\ filament. 
At the eastern boundary of the remnant,
we see a faint CO filament that appears to be spatially coincident 
with the radio/\hh\ filament, but the CO emission is complex in this area
and their association is not clear.

To summarize, we could not identify CO clouds 
clearly associated with the \hh\ filaments (and the SNR). 
The large molecular shell at $v_{\rm LSR}=+51$--$+55~\kms$ has been 
proposed to be associated with the SNR \citep{paron13},
but its velocity is very different from that of the \hh\ filaments, and 
there is no spatial correlation between the two.
Instead, we have detected diffuse CO emission features 
spatially coincident with the SNR 
at the velocities of the \hh\ filaments ($+72$--$+85~\kms$).
Their association is possible, although there is no clear spatial correlation 
between the CO and \hh\ emission features.
The kinematic distance corresponding to 
the \hh\ emission-line velocities (+73--+85~\kms) is 
5.0--5.5 kpc (Table~\ref{tab-dis}) assuming the near-side distance.

\subsection{G18.9$-$1.1} \label{sec-dis-g18.9}
G18.9$-$1.1 belongs to the composite-type SNR
with a size larger than a half degree.
Early radio observations showed 
weak, diffuse radio continuum filling inside the remnant,
together with two strong pillar-like features
at the center and near the western border \citep[e.g.,][]{fur85,fur89,fur97}.
X-ray observations found
a point-like hard X-ray source with an associated diffuse nebula 
at the tip of the central radio pillar,
and it was suggested that these objects are 
a pulsar and its wind nebula \citep{har04,tul10}. 
\citet{fur89} reported the detection of \hi\ depression at $+18~\kms$
toward the remnant and suggested that the depression is caused by a cavity 
associated with the remnant (see below).
The $v_{\rm LSR}$ of $+18~\kms$ corresponds to the kinematic distance of
either $\sim2$ kpc within the Sagittarius Arm
or $\sim15$ kpc on the far side of the tangential point.
The latter, however, was ruled out 
because in that case the SNR should have very unusual properties
\citep[e.g, huge explosion energy of
($1$--$2$)$\times 10^{52}$ erg;][]{fur89,har04}.
A recent NIR study using the red clump stars toward the remnant
suggested the distance of $1.8\pm0.2$~kpc \citep{shan18},
which is consistent with the above kinematic distance.

In \citet{lee19},
we detected weak, patchy \hh\ emission inside the remnant.
Most of the \hh\ emission, which has a complex morphology,
arises from the central and northeastern regions,
$3\arcmin$--$10\arcmin$ apart from the central radio pillar.
Additional weak, patchy \hh\ emission features were also detected
in the western border of the remnant.
We obtained two NIR spectra:
one at the \hh\ filament close to the central radio pillar, and
the other at the \hh\ clump near the northeastern border
(Figure~\ref{fig-slitpos-1}).
The central velocity of the latter is +70~\kms, while the spectrum of the 
former is composed of two velocity components
centered at +74 and +53~\kms, respectively
(Table~\ref{tab-prop}; see also Figure~\ref{fig-spec}).
If we adopt the central velocity of the northeastern \hh\ clump
as the systemic velocity of the SNR,
the kinematic distance to the SNR would be 4.7 kpc (Table~\ref{tab-dis}).
But this $v_{\rm LSR}$ is more than $50~\kms$ larger than
that ($+18~\kms$) suggested by \citet{fur89},
and the distance is $\sim 3$~kpc larger than
those ($\sim 2$~kpc) obtained in the previous radio and NIR studies
\citep{fur89,shan18}.

The large velocity difference in $v_{\rm LSR}$
between our \hh\ emission and the previous \hi\ absorption could be due 
to the peculiar velocity of the \hh\ gas accelerated by the SN shock
along the line of sight.
However, the systemic velocity ($v_{\rm LSR}=+70$~\kms) is taken from
the northeastern \hh\ clump near the SNR boundary,
so the peculiar line-of-sight velocity by the SN shock might not be large.
Furthermore, the velocity difference of $50~\kms$ seems to be 
rather large for a nondissociative shock 
\citep[e.g.,][]{draine93}.
Therefore, it is possible that
the $v_{\rm LSR}$ of the \hh\ emission lines
traces the systematic velocity of the remnant.
Another possibility is that
these central and northeastern \hh\ emission features
are not physically associated with
the SNR, even though they are located inside the remnant
(Figure~\ref{fig-slitpos-1}).
We searched the WISE Catalog of Galactic \hii\ Regions
\citep[V2.2;][]{anderson14}
and the HASH Planetary Nebulae Database\footnote{
The University of Hong Kong/Australian Astronomical Observatory/Strasbourg
Observatory H$\alpha$ Planetary Nebula Database}
\citep[V4.6;][]{parker16}
to find sources responsible for the \hh\ emission,
but we could not find any candidate.

In order to explore the large discrepancies in $v_{\rm LSR}$ and
distance between our \hh\ and previous results,
we examined the \hi\ absorption toward the SNR using
the VLA Galactic Plane Survey (VGPS) \hi\ data \citep{sti06}.
The VGPS survey has an angular resolution of $1\arcmin$,
which is much higher than that ($6\arcmin$) of the previous study
\citep{fur89}. 
We first extracted \hi\ spectra toward the two central radio pillars 
and compared them with the spectra of the surrounding regions,
but we could not identify clear absorption features because
of the random fluctuations in the foreground/background \hi\ emission.
We then looked at individual channel maps and 
confirmed the depression at $v_{\rm LSR} \sim +18$~\kms~(Figure~\ref{fig-g189}).
The depression has a very good morphological correlation with  
radio continuum,
e.g., there is a filamentary depression along the radio pillars, 
suggesting that the depression is probably due to absorption 
rather than a cavity as proposed by \cite{fur89}.
We could not find absorption features at higher velocities, but the 
nondetection of absorption does not rule out the possibility that the SNR 
is at a greater distance because there is a large background fluctuation 
due to the nonuniform structure of the \hi\ gas along the line of sight. 
The \hi\ absorption at $+18~\kms$ indicates that the SNR is at a distance 
{\it greater} than $\sim 2$~kpc.
The distance $1.8\pm0.2$~kpc of \citet{shan18} has been obtained by
comparing the X-ray absorbing hydrogen column density
\citep[$(8.3 \pm 0.5) \times 10^{21}$ cm$^{-2}$;][]{har04}
to the extinction-distance relation derived by using red clump stars.  
\citet{har04}, however, obtained an extinction $A_{V}=5.1$ mag,
which is 1 mag greater than the value of 4.1 mag from \citet{shan18}.
And the extinction-distance relation of \citet{shan18} shows that 
$A_{V}$ is almost constant at $\sim 5.2$~mag 
beyond 2 kpc. (They derived the relation up to 2.8 kpc.) 
Therefore, we consider that
the distance 2 kpc obtained in previous studies is a minimum distance,
and the distance to G18.9$-$1.1 is still uncertain.
Future spectroscopic studies of the \hh\ emission features 
detected in the western radio shell \citep{lee19}
might be helpful to clarify the issue.

\subsection{G21.8$-$0.6 (Kes 69)} \label{sec-dis-kes69}
Kes 69 is a shell-type SNR, showing a bright incomplete radio shell
along the southeastern border of the remnant.
\hi\ absorption is seen up to $+85~\kms$ with no absorption lines
around the tangential velocity ($\sim +110~\kms$), 
indicating that the remnant is located on the near side of the tangential point
\citep{tia08,zho09}.
The kinematic distance of the SNR corresponding to $+85~\kms$ is
5.2 kpc \citep[see][]{ran18b}.
\citet{zho09} showed that a molecular arc at $+77$--$+86~\kms$ 
is morphologically correlated with the southeastern radio shell. 
\citet{hew08} detected extended faint OH 1720 MHz maser emission  
at $v_{\rm LSR}=+85~\kms$ toward the brightest portion of
the southeastern radio shell.
A compact OH (1720 MHz) maser spot was also detected
in the northeastern region of the remnant,
but at $v_{\rm LSR}=+69.3~\kms$ \citep{gre97}.

In \citet{lee19},
we reported the detection of an extended bright  \hh\ emission feature
composed of multiple narrow filaments 
in the southeastern border of the remnant along the bright radio shell.
We also detected additional complex \hh\ emission features 
in the northeastern region of the remnant,
where the OH maser has been detected \citep{gre97}.
We took NIR spectra at three positions:
two in the southeastern shell and one in the northeastern region
(Table~\ref{tab-prop}).
The $v_{\rm LSR}$ of the \hh\ emission in the southeastern shell is $+60~\kms$,
and it is consistent with the previous NIR spectroscopy with medium resolution
for the southeastern shell \citep[$+57\pm3~\kms$;][]{lee19}.
Interestingly, this velocity is almost $30~\kms$ smaller than
the $v_{\rm LSR}$ of the molecular arc detected in CO and
OH maser emission \citep[$+85~\kms$;][]{hew08,zho09}.
The large velocity difference is difficult to understand
because the \hh\ filaments are located at the border of the remnant,
so that the expansion velocity of \hh\ gas 
along the line of sight might be small.

We have investigated the presence of MCs associated with the remnant at 
the velocity of the \hh\ emission using the FUGIN $^{12}$CO $J=1$--$0$ data. 
Figure~\ref{fig-kes69} shows CO channel maps
at velocities from +57 to +93~\kms,
where we see that the most prominent MCs are at $+78$--$+88~\kms$.
As shown by \citet{zho09}, the MCs in this velocity range
have some morphological correlation with the SNR.
In particular, the filamentary MC in the southeastern area at $\sim +80$~\kms\ 
correlates well with the remnant's bright southeastern radio continuum shell.
There is also extended OH maser emission with a broad line width 
at +85~\kms along the radio shell \citep{hew08}.
At lower velocities, the CO emission is relatively faint 
with some large, extended CO clouds inside and around the SNR. 
At the velocity of the \hh\ emission ($\sim +61~\kms$),
we can see a faint, clumpy MC that appears to be in contact with 
the southern SNR boundary (Figure~\ref{fig-kes69_zoom}). 
It has a thin, extended filamentary structure 
parallel to the \hh\ filament in the southeastern shell, 
but the overall morphological correlation 
between the MC and the radio shell is relatively weak.

If the MCs at $+78$--$+88~\kms$ in Figure~\ref{fig-kes69}
are associated with the SNR,
as has been suggested in previous molecular line studies, 
the kinematic distance to Kes 69 is 5.2~kpc,  
adopting +85~\kms\ as its systemic velocity \citep{hew08,zho09}.
The \hh\ emission with a central velocity of +61~\kms, then 
indicates that the MC is on the front side of the SNR and that   
a $\sim 24$~\kms\ shock is propagating into the MC.  
On the other hand, if the central velocity of the \hh\ emission (+61~\kms) 
represents the systemic velocity of the SNR, 
the faint CO cloud in Figure~\ref{fig-kes69_zoom} is
possibly interacting with the SNR, and the association of
the extended OH maser emission detected by \citet{hew08} becomes unclear.
The kinematic distance corresponding to 
the systemic velocity of $+61~\kms$ is 4.1~kpc.
Future spectral mapping of the \hh\ emission might be helpful 
to resolve the issue.

\subsection{G39.2$-$0.3 (3C 396)} \label{sec-dis-3c396}
3C 396 is a composite-type SNR exhibiting both a central X-ray
pulsar wind nebula (PWN)
and an incomplete radio shell composed of multiple filaments \citep{har99}.
Previous \hi\ and OH absorption studies suggested that
the remnant is located on the far side of the tangential point 
at $v_{\rm LSR} \lesssim +60~\kms$
(\citealt{cas75}; see also \citealt{gre89}).
The remnant is believed to be interacting with MCs in the west,
where the bright radio shell and NIR \hh\ filaments are detected.
According to CO observations,
there are two large MCs that could be possibly associated with the remnant:
one at $+69~\kms$ and the other at $+84~\kms$.
\citet{lee09} found MCs at $v_{\rm LSR}=+68$--$+70~\kms$
surrounding the remnant
and pointed out that the western \hh\ filament is in contact with
the inner boundary of the MCs. 
\citet{kil16} also suggested the association of these MCs with the SNR.
On the other hand, \citet{su11} 
found a thick molecular wall at $+84~\kms$ in the western border of the remnant
with high $^{12}$CO $J$=2--1/$J$=1--0 line ratios
and suggested that the SNR is interacting with these MCs.

Previous NIR \hh\ narrowband imaging observations had detected
two long \hh\ filaments aligned along the north-south direction,
leaving an interval of $\sim 1\farcm5$
in the western region of the remnant \citep{lee09,lee19}.
The brighter filament is located near the western edge of the remnant,
slightly outside the radio boundary of the SNR.
\citet{lee09} attributed the \hh\ filaments
to the interaction between the SNR and nearby MCs at $+69~\kms$.
Our NIR spectroscopy toward the two locations in the filament
shows that $v_{\rm LSR}$ of the filament is $+56~\kms$ (Table~\ref{tab-prop}).
This is consistent with
the result of an early \hi\ absorption study \citep{cas75},
but it is considerably (by $10$--$20~\kms$) smaller than those inferred from 
previous CO observations \citep[$+69$, $+84~\kms$;][]{lee09,su11,kil16}.

In Figure~\ref{fig-3c396},
we revisit the $^{12}$CO $J=1$--$0$ channel maps from $+52$ to $+91~\kms$
to find the molecular features associated with the remnant.
The most prominent feature are the large MCs at velocities 
from $+52$ to $+70~\kms$ in the northern and western regions of the remnant.
At $+52~\kms$, these large MCs appear in the western region, 
$2\arcmin$--$4\arcmin$ apart from the remnant boundary. 
They move to the central area of the field at higher velocities, so that 
at $\sim +57~\kms$ they are almost in contact with the remnant boundary.
Around $+54$--$+57~\kms$,
we also see extended MCs in the eastern region of the remnant,
which seems to be in contact with the eastern boundary of the remnant.
From $+62$ to $+70~\kms$, large MCs appear in the northern region.
A couple of MC filaments extend from the northern MCs toward the south,
and one of them at $+70~\kms$ is almost in contact with
the SNR along the southwestern boundary of the remnant.
\citet{lee09} suggested that
this MC filament is physically associated with the remnant.
At velocities $>+72~\kms$,
we see relatively diffuse, extended MCs around the remnant,
which disappear at velocities 
higher than the tangential velocity ($\sim+85~\kms$).
In particular, at +85~\kms, 
we see several clouds superposed on the bright western radio shell.
\citet{su11} found high $^{12}$CO $J$=2--1/$J$=1--0 line ratios
at the velocity channel
and suggested that these CO clouds are interacting with the remnant.

Figure~\ref{fig-3c396_zoom} shows
the zoomed-in $^{12}$CO channel map of 3C 396 at $+56~\kms$,
which is the central velocity of the \hh\ emission lines.
An interesting thing is that the remnant is located 
inside a cavity surrounded by MCs.
The western half of the cavity is prominent, and its 
inner boundary appears to be parallel to the SNR boundary.  
Except the northwestern region,
there is a $\sim 2\arcmin$ gap between the 
remnant boundary and the cavity wall.
But there is a filamentary cloud 
spatially coincident with the \hh\ filament inside the cavity. 
The eastern half of the cavity is surrounded 
by relatively faint MCs. The SNR radio emission is 
faint in this area, but the SNR appears to be 
in contact with the cavity wall along the southeastern boundary of the SNR.

As we have shown in Figure~\ref{fig-3c396},
there are at least three possible MCs
at velocities +56, +69, and +84~\kms, respectively,
that could be associated with the SNR \citep[see also][]{lee09,su11,kil16}.
They all have some morphological relation with the SNR. 
At the velocity of the \hh\ emission, i.e., at +56~\kms,
we see a CO cavity surrounding the SNR 
and a faint filamentary MC spatially coincident with the bright western shell.
Therefore, the MCs at +56~\kms\ are very likely to be associated with the SNR.  
The cavity surrounding the SNR could be a wind bubble
produced by the progenitor. 
The SNR is on the far side \citep{cas75}, so that 
the kinematic distance corresponding to +56~\kms\ is 9.5 kpc.

\subsection{G43.3$-$0.2 (W49B)} \label{sec-dis-w49b}
W49B is a member of the W49 complex that 
includes multiple \hii\ regions (W49A) and one SNR (W49B).
Previous \hi\ 21~cm observations toward the W49 complex showed that 
the absorption is seen up to the tangential velocity ($\sim+70~\kms$)
at positive velocities,
but not at negative velocities \citep[][]{loc78,bro01,zhu14,ran18a}.
This indicates that the sources in the W49 complex 
are located beyond the tangential point but inside the solar circle.
There have been many efforts to reveal the relative 
distances to W49A and W49B and the physical association between them.
In \hi\ and H$_{2}$CO absorption spectra, it was found that 
there is an absorption at $\sim+10~\kms$
toward W49A that is not seen toward W49B \citep{kaz70,wil70,rad72}.
This suggests that W49A is most likely located at the far distance 
(11.5 kpc), corresponding to $+10~\kms$ and that 
W49B is in front of W49A without any physical association between them 
\citep{kaz70,wil70,rad72}.
This distance to W49A was later confirmed by
the comparison of proper motions of the H$_{2}$CO masers
and their Doppler velocities
\citep[$11.4\pm1.2$ kpc;][]{gwi92}.
The early \hi\ and H$_{2}$CO absorption studies suggested that
W49B is located at $\sim 8$~kpc \citep{moffett94}
corresponding to the maximum velocity of the absorption lines
\citep[$+63~\kms$;][]{kaz70,wil70,rad72}.
\citet{bro01}, however, suggested that
the different absorption velocities between W49A and W49B 
could be due to
the differences of \hi\ kinematics, distribution, and temperature
in the direction of W49 complex 
and that one cannot rule out the physical association between them.
Another distance estimate for W49B is from CO observations.
Toward W49B, there are three large MCs at different $v_{\rm LSR}$:
$+10$, $+40$, and $+60~\kms$ \citep{zhu14,kil16}.
\citet{zhu14} argued that
the MCs at $+40~\kms$ are morphologically associated with the remnant
and suggested a distance of $\sim10$ kpc.
\citet{kil16}, on the other hand, reported
the detection of broadened $^{12}$CO emission centered at
$v_{\rm LSR}=+14~\kms$ in the southwestern boundary of the remnant
and suggested 11.3 kpc, although 
the physical association between the remnant and the MCs 
is not clear.

In \citet{lee19},
we found bright \hh\ emission features
in the eastern and western areas of the remnant.
In the eastern area,
we see several narrow, extended filaments along the NE-SW direction 
outside the X-ray/\feii/radio boundary.
In the western area, on the other hand,
the \hh\ emission features are rather clumpy
and mostly detected inside the remnant.
There are also some additional \hh\ filaments aligned along the NE-SW direction
in the central area of the remnant,
but they are much fainter than those in the eastern and western areas.
The morphology and the \hh\ line ratios (Section~\ref{sec-res-ext}) 
indicate that the bright \hh\ emission features in the east and west areas 
(and probably the faint filaments in the central area too) 
are thermally excited and physically associated with the remnant.
We obtained NIR spectra at four positions:
two for the eastern filaments,
and one for each of the northern and the western \hh\ emission features
(Figure~\ref{fig-slitpos-2}).
We found that their $v_{\rm LSR}$ are $\sim +64~\kms$
without any significant variations (only $\pm2~\kms$),
and this led us to conclude that
the systematic velocity of the remnant is $+64~\kms$.
Our result is consistent with the suggestions by
the early \hi\ and H$_{2}$CO absorption studies \citep{kaz70,wil70,rad72},
but is very different from
those of previous CO observations, i.e., $+10$ or $+40~\kms$
\citep{zhu14,kil16}.
The kinematic distance corresponding to $+64~\kms$ is
7.5 kpc (Table~\ref{tab-dis}).

We revisit $^{12}$CO channel maps
to confirm the MC(s) associated with the SNR. 
Figure~\ref{fig-w49b} shows the $^{12}$CO $J=1$--$0$ channel maps
toward W49B in three different velocity ranges
where the large MCs are located \citep{zhu14,kil16}:
(1) $+7$--$+17~\kms$, (2) $+38$--$+48~\kms$, and (3) $+59$--$+69~\kms$.
In the velocity range from $+7$ to $+17~\kms$,
there are two large prominent MCs, i.e.,  
an extended filamentary MC in the northern area of the field and  
a large MC outside the western boundary of the SNR.  
There is diffuse CO emission spread over the SNR 
that appears to be extended from these two MCs.
There is no clear overall morphological correlation
between the diffuse emission and the \hh\ emission.
But at $+14~\kms$, there is a $\sim 4\arcmin$-long filamentary CO cloud
along the southwestern boundary of the SNR
that appears to surround the \hh-bright area. 
\citet{kil16} found broadened CO emission lines toward this cloud 
suggesting the association of the MC with the SNR. 
The presence of the \hh\ emission further supports the SNR-MC interaction,
but the large ($\sim 50~\kms$) velocity difference
between the \hh\ and the CO emission is difficult to reconcile.
In the velocity range from $+38~\kms$ to $+48~\kms$,
there are diffuse CO clouds around the remnant.
As pointed out by \citet{zhu14}, the MCs appear to surround the SNR 
(see also Figure~1 of \citealt{chen14}).
In particular, at $+43~\kms$,
there are MCs just outside the eastern and southwestern \hh-bright regions. 
If these MCs are interacting with the SNR,
the velocity of the \hh\ emission ($+64~\kms$) 
implies that the MCs are on the backside of the SNR.
In the velocity range $+59$--$+69$~\kms, 
the CO emission is complex with several velocity components.
Note that this is the velocity range
where \hi\ and H$_2$CO absorption is prominent,
especially toward the southwestern area of the SNR
(\citealt{bieging82,bro01}; see also \citealt{lacey01}).
Toward the southwestern \hh-bright region,
there are at least two velocity components
at $+61$ and $+66~\kms$, respectively.
The $+61~\kms$ component is part of
the large MC aligned along the northeast-southwest direction
crossing the remnant's center.
At this velocity,
we also see a relatively faint CO filament outside the eastern SNR boundary 
that appears to be aligned with the eastern \hh-bright region
(see next paragraph). 
The $+66~\kms$ component seems to be part of the large 
MC cloud in the western area of the field,
which partly overlaps with the southwestern \hh-bright region. 
There is also some faint CO emission at this velocity
just outside the eastern \hh-bright region, but only in its southern area.

Figure~\ref{fig-w49b_zoom} is a zoomed-in $^{12}$CO channel map
integrated from $+61$ to $+66~\kms$,
where we found the \hh\ emission lines.
Note that the color scale is changed from that in Figure~\ref{fig-w49b}.
The figure shows that the $+61~\kms$ MC crossing the SNR 
has a ``tuning-fork'' morphology with the U-shaped prongs in the interior.   
The left prong runs along the northeast-southwest direction,
spatially coincident with one of the radio filaments in the interior,
while the western prong is located just inside the 
western radio shell, where the radio continuum is bright and the 
\hh\ emission lines are detected. 
In the eastern area, along the extended \hh-bright region, 
there are faint MCs that appear, blocking the \hh\ region.
These MCs are aligned with the tuning-fork MC but with a gap between them. 
There is also diffuse emission 
spread over the field, including the areas around
the southwestern \hh-bright region.

As we have shown in Figure~\ref{fig-w49b},
the CO emission toward W49B is complicated, and it is difficult 
to identify MCs associated with the SNR based on their morphology. 
Instead, the roughly constant central velocities of the \hh\ emission 
over the SNR suggest that the systemic velocity of the 
SNR is probably $\sim+64~\kms$.
We therefore consider that the MCs at $+59$--$+69~\kms$ 
are associated with the SNR. 
In particular, the MCs at $+61$--$+66~\kms$
seem to have a good morphological correlation with 
the SNR (Figure~\ref{fig-w49b_zoom}).
The interaction of the SNR with these 
MCs aligned along the northeast-southwest direction
could have resulted in the barrel-like morphology of the SNR.
The kinematic distance 7.5 kpc corresponding to +64~\kms\ is
$\sim4$ kpc closer than W49A \citep[$11.4\pm1.2$ kpc;][]{gwi92}.

\section{Summary} \label{sec-sum}
Distance is a basic parameter of the Galactic SNRs
and is essential for deriving their physical parameters, such as 
radius, luminosity, age, and explosion energy.
Diverse methods have been devised, each of which has its own 
strengths and weaknesses. 
A popular method is to identify 
an object associated with the SNR and derive  
the kinematic distance from its LSR velocity.
This technique has been most widely used in radio 
by using the CO emission from MCs or 
OH maser emission. 
In this paper, we have shown that the NIR H$_2$ emission, which is a  
strong signature of the interaction between SNRs and MCs,
can also be used for the distance determination.
We performed high-resolution NIR spectroscopy for 
16 SNRs detected in our systematic \hh\ emission-line study of 
SNRs in the inner Galaxy \citep{lee19}.
In all SNRs, bright \hh\ 1--0 S(1) 2.122~\micron\ emission lines,
together with additional \hh\ emission lines from different excitation levels,
were clearly detected.
We examined the excitation mechanism of the \hh\ emission lines
in order to confirm their physical association with the SNRs,
and we derived the kinematic distances of SNRs 
from the central velocities of the \hh\ lines.
For four SNRs,
the kinematic distances have been determined in this work for the first time.

The comparison showed that
the central velocities of \hh\ emission lines are considerably ($>5~\kms$)
different from those obtained from radio observations for fives SNRs:
G18.1$-$0.1, G18.9$-$1.1, Kes 69, 3C 396, and W49B.
Since the detection of collisionally excited \hh\ emission in an SNR is
strong evidence that the SNR is interacting with an MC,
we consider that the \hh\ central velocities and the inferred kinematic distances
should be preferred to those based on morphological association
if there is a large discrepancy.
However, as pointed out in Section~\ref{sec-res-dis},
the kinematic distance from the \hh\ line,
with the matter of their physical association put aside,
also has its inherent uncertainties.
Firstly, the measured velocity is that of the shocked gas, not the ambient gas.
Secondly, the \hh\ emission samples a highly localized region within an MC,
so that the velocity may represent the outskirts of cloud velocity distribution
rather than the central velocity of the cloud.
More detailed studies are needed to fully accept
the \hh\ line-based distances for these five SNRs.

Our main results are summarized in the following:

\noindent
1. In all SNRs, except G9.9$-$0.8, 
the \hh\ lines show thermal line ratios with $T \sim 2000$ K,
indicating that they are emitted from 
warm, collisionally excited \hh\ gas.
The thermal line ratios support that most, if not all, 
\hh\ emission features are associated with the SNRs.

\noindent
2. We have derived the kinematic distances of 16 SNRs 
(including  G9.9$-$0.8)
by using the velocities of the \hh\ emission lines 
(see below for the discussion about G9.9$-$0.8). 
For four SNRs (G13.5$+$0.2, G16.0$-$0.5, G32.1$-$0.9, and G33.2$-$0.6),
the kinematic distances have been determined in this work for the first time.
For 6 out of the remaining 11 SNRs,
$v_{\rm LSR}$ derived from our \hh\ emission lines are
well consistent ($<\pm5~\kms$) with those from previous radio observations,
which provides a support to the derived kinematic distances.
For five SNRs, however, there is a large 
discrepancy between the two velocities, which 
casts doubt on the derived kinematic distances.

\noindent
3. For five SNRs showing large discrepancy  
in $v_{\rm LSR}$ between our \hh\ and the previous radio observations
(G18.1$-$0.1, G18.9$-$1.1, Kes 69, 3C 396, W49B),
we have explored the origin of the discrepancy using CO and \hi\ data.
The results on individual SNRs are summarized below.  

(1) G18.1$-$0.1: The velocities of the two extended \hh\ filaments
along the northern and eastern radio boundaries 
have been measured as $+73$ and $+85~\kms$, respectively.
These velocities are very different from those suggested from  
previous CO \citep[$+51$--$+55~\kms$;][]{paron13} and
\hi\ \citep[$+100~\kms$;][]{lea14,ran18a} observations. 
At the velocities of the \hh\ emission,
there are some diffuse CO emission features 
spatially coincident with the remnant.
Their association is possible.
The near-side kinematic distance
corresponding to the \hh\ emission line velocities is 5.0--5.5 kpc.

(2) G18.9$-$1.1: The velocities of two short \hh\ filaments have been measured:
one near the northeastern boundary ($+70~\kms$), and
the other in the central area
($+74$ and $+53~\kms$; see Figure~\ref{fig-spec}).
The near-side kinematic distance corresponding to $+70~\kms$ is 4.7 kpc.
These are much greater than those suggested in
previous \hi\ \citep[$+18~\kms$;][]{fur89} and
red clump star \citep[1.8~kpc][]{shan18} studies.
The discrepancy may indicate that either
the \hh\ filaments have a large peculiar velocity or
they are not associated with the SNR.
The ISM structure along this line of sight, however, appears complicated,
and the previous velocity/distance estimates could be lower limits.  

(3) Kes 69: The velocity of the \hh\ filament 
in the southeastern bright radio shell has been measured as $+60~\kms$.
This is much smaller than the value of $+85~\kms$ of the molecular arc
suggested to be interacting with the SNR from CO and OH maser emission
\citep[$+85~\kms$;][]{hew08,tia08,zho09}.
At the velocity of the \hh\ emission, however,
we see a faint, clumpy MC that could be responsible for the \hh\ emission.
If this $+61~\kms$ MC is interacting with the SNR,  
the near-side kinematic distance becomes 4.1~kpc.

(4) 3C 396: The velocity of the bright \hh\ filament coincident with
the western radio shell has been measured as $+57~\kms$.
This velocity is considerably smaller than
those suggested from previous CO observations,
i.e., $+69~\kms$ \citep{lee09,kil16} and $+84\kms$ \citep{su11}.
We have found that, at $+56~\kms$,
there are MCs spatially well correlated with the SNR:
a CO cavity surrounding the SNR and a faint filamentary MC 
spatially coincident with the bright western shell. 
If the MCs at $+56~\kms$ are associated with the SNR,
the kinematic distance to 3C 396, which is on the far side \citep{cas75},
becomes 9.5 kpc. 

(5) W49B: The velocities of the bright \hh\ filaments in the 
the eastern and western SNR shells 
have been measured as $\sim +64~\kms$.
This velocity is very different from those suggested from CO observations, 
i.e., $+40~\kms$ \citep{zhu14} and $+14~\kms$ \citep{kil16}.
The roughly constant velocity of the \hh\ emission over the SNR suggests that
the systemic velocity of the SNR is probably $\sim +64~\kms$.
At $+61$--$+66~\kms$, we have found MCs
that appear to have a good morphological correlation with
the barrel-like morphology of the SNR.
Adopting $+64~\kms$ as the systemic velocity of the SNR, 
the kinematic distance to W49B, which is on the far side
\citep{loc78,bro01,zhu14,ran18a}, becomes 7.5 kpc.

\noindent
4. For G9.9$-$0.8,
the nonthermal lines ratios and the narrow line widths indicate that
the \hh\ lines are probably excited by UV radiation. 
The \hh\ emission is most likely arising from a PDR
surrounding the \hii\ region (G9.982$-$0.752), rather than the SNR.
However, the central velocity of the \hh\ line ($+30~\kms$) is
almost identical to the systematic velocity of the remnant
estimated from previous CO observations \citep[$+31~\kms$;][]{kil16},
suggesting the possible physical association
between the \hii\ region and the SNR.

\acknowledgments
Y.-H.L. acknowledges support from the National Research Foundation of Korea (NRF)
grant funded by the Ministry of Science and ICT (MSIT) of Korea
(NRF-2018M1A3A3A02065645).
The work of B.-C.K. was supported by the Basic Science 
Research Program through the National Research Foundation of
Korea (NRF) funded by the Ministry of Science, ICT and future Planning
(2020R1A2B5B01001994).
This work used the Immersion Grating Infrared Spectrograph (IGRINS)
that was developed under a collaboration between
the University of Texas at Austin and
the Korea Astronomy and Space Science Institute (KASI)
with the financial support of
the US National Science Foundation under grant AST-1229522,
the University of Texas at Austin,
and the Korean GMT Project of KASI.
This paper includes data taken at
the McDonald Observatory of the University of Texas at Austin.
This publication also makes use of data from FUGIN,
the FOREST Unbiased Galactic plane Imaging survey
with the Nobeyama 45 m telescope,
a legacy project in the Nobeyama 45 m radio telescope.
The 45 m radio telescope is operated by the Nobeyama Radio Observatory,
a branch of the National Astronomical Observatory of Japan.


\bibliographystyle{yahapj}
\bibliography{references}

\begin{thebibliography}{}
\providecommand\natexlab[1]{#1}
\providecommand\JournalTitle[1]{#1}

\bibitem[{{Anderson} {et~al.}(2014){Anderson}, {Bania}, {Balser}, {Cunningham},
  {Wenger}, {Johnstone}, \& {Armentrout}}]{anderson14}
{Anderson}, L.~D., {Bania}, T.~M., {Balser}, D.~S., {et~al.} 2014,
  \href{http://dx.doi.org/10.1088/0067-0049/212/1/1}{\JournalTitle{\apjs}, 212,
  1}

\bibitem[{{Bieging} {et~al.}(1982){Bieging}, {Wilson}, \& {Downes}}]{bieging82}
{Bieging}, J.~H., {Wilson}, T.~L., \& {Downes}, D. 1982, \JournalTitle{\aaps},
  49, 607

\bibitem[{{Black} \& {van Dishoeck}(1987)}]{bla87}
{Black}, J.~H., \& {van Dishoeck}, E.~F. 1987,
  \href{http://dx.doi.org/10.1086/165740}{\JournalTitle{\apj}, 322, 412}

\bibitem[{{Brogan} {et~al.}(2006){Brogan}, {Gelfand}, {Gaensler}, {Kassim}, \&
  {Lazio}}]{bro06}
{Brogan}, C.~L., {Gelfand}, J.~D., {Gaensler}, B.~M., {Kassim}, N.~E., \&
  {Lazio}, T.~J.~W. 2006,
  \href{http://dx.doi.org/10.1086/501500}{\JournalTitle{\apjl}, 639, L25}

\bibitem[{{Brogan} \& {Troland}(2001)}]{bro01}
{Brogan}, C.~L., \& {Troland}, T.~H. 2001,
  \href{http://dx.doi.org/10.1086/319787}{\JournalTitle{\apj}, 550, 799}

\bibitem[{{Burton} \& {Spyromilio}(1993)}]{burton93}
{Burton}, M., \& {Spyromilio}, J. 1993, \JournalTitle{Proceedings of the
  Astronomical Society of Australia}, 10, 327

\bibitem[{{Burton}(1992)}]{bur92}
{Burton}, M.~G. 1992,
  \href{http://dx.doi.org/10.1071/PH920463}{\JournalTitle{Australian Journal of
  Physics}, 45, 463}

\bibitem[{{Burton} {et~al.}(1989){Burton}, {Brand}, {Geballe}, \&
  {Webster}}]{burton89}
{Burton}, M.~G., {Brand}, P.~W.~J.~L., {Geballe}, T.~R., \& {Webster}, A.~S.
  1989, \href{http://dx.doi.org/10.1093/mnras/236.3.409}{\JournalTitle{\mnras},
  236, 409}

\bibitem[{{Burton} {et~al.}(1988){Burton}, {Geballe}, {Brand}, \&
  {Webster}}]{burton88}
{Burton}, M.~G., {Geballe}, T.~R., {Brand}, P.~W.~J.~L., \& {Webster}, A.~S.
  1988, \href{http://dx.doi.org/10.1093/mnras/231.3.617}{\JournalTitle{\mnras},
  231, 617}

\bibitem[{{Caswell} {et~al.}(1975){Caswell}, {Murray}, {Roger}, {Cole}, \&
  {Cooke}}]{cas75}
{Caswell}, J.~L., {Murray}, J.~D., {Roger}, R.~S., {Cole}, D.~J., \& {Cooke},
  D.~J. 1975, \JournalTitle{\aap}, 45, 239

\bibitem[{{Chen} {et~al.}(2014){Chen}, {Jiang}, {Zhou}, {Su}, {Zhou}, {Li}, \&
  {Zhang}}]{chen14}
{Chen}, Y., {Jiang}, B., {Zhou}, P., {et~al.} 2014,
  \href{http://dx.doi.org/10.1017/S1743921313009423}{in IAU Symposium, Vol.
  296, Supernova Environmental Impacts, ed. A.~{Ray} \& R.~A. {McCray}}, 170

\bibitem[{{Clark} {et~al.}(1962){Clark}, {Radhakrishnan}, \& {Wilson}}]{cla62}
{Clark}, B.~G., {Radhakrishnan}, V., \& {Wilson}, R.~W. 1962,
  \href{http://dx.doi.org/10.1086/147255}{\JournalTitle{\apj}, 135, 151}

\bibitem[{{Claussen} {et~al.}(1997){Claussen}, {Frail}, {Goss}, \&
  {Gaume}}]{cla97}
{Claussen}, M.~J., {Frail}, D.~A., {Goss}, W.~M., \& {Gaume}, R.~A. 1997,
  \href{http://dx.doi.org/10.1086/304784}{\JournalTitle{\apj}, 489, 143}

\bibitem[{{Clemens}(1985)}]{clemens85}
{Clemens}, D.~P. 1985,
  \href{http://dx.doi.org/10.1086/163386}{\JournalTitle{\apj}, 295, 422}

\bibitem[{{Draine}(2003)}]{dra03}
{Draine}, B.~T. 2003,
  \href{http://dx.doi.org/10.1146/annurev.astro.41.011802.094840}{\JournalTitle{\araa},
  41, 241}

\bibitem[{{Draine} \& {Bertoldi}(1996)}]{dra96}
{Draine}, B.~T., \& {Bertoldi}, F. 1996,
  \href{http://dx.doi.org/10.1086/177689}{\JournalTitle{\apj}, 468, 269}

\bibitem[{{Draine} \& {McKee}(1993)}]{draine93}
{Draine}, B.~T., \& {McKee}, C.~F. 1993,
  \href{http://dx.doi.org/10.1146/annurev.aa.31.090193.002105}{\JournalTitle{\araa},
  31, 373}

\bibitem[{{Draine} \& {Woods}(1990)}]{dra90}
{Draine}, B.~T., \& {Woods}, D.~T. 1990,
  \href{http://dx.doi.org/10.1086/169358}{\JournalTitle{\apj}, 363, 464}

\bibitem[{{Folgheraiter} {et~al.}(1997){Folgheraiter}, {Warwick}, {Watson}, \&
  {Koyama}}]{fol97}
{Folgheraiter}, E.~L., {Warwick}, R.~S., {Watson}, M.~G., \& {Koyama}, K. 1997,
  \href{http://dx.doi.org/10.1093/mnras/292.2.365}{\JournalTitle{\mnras}, 292,
  365}

\bibitem[{{Frail} {et~al.}(1996){Frail}, {Goss}, {Reynoso}, {Giacani}, {Green},
  \& {Otrupcek}}]{fra96}
{Frail}, D.~A., {Goss}, W.~M., {Reynoso}, E.~M., {et~al.} 1996,
  \href{http://dx.doi.org/10.1086/117904}{\JournalTitle{\aj}, 111, 1651}

\bibitem[{{Froebrich} {et~al.}(2011){Froebrich}, {Davis}, {Ioannidis},
  {Gledhill}, {Takami}, {Chrysostomou}, {Drew}, {Eisl{\"o}ffel}, {Gosling},
  {Gredel}, {Hatchell}, {Hodapp}, {Kumar}, {Lucas}, {Matthews}, {Rawlings},
  {Smith}, {Stecklum}, {Varricatt}, {Lee}, {Teixeira}, {Aspin}, {Khanzadyan},
  {Karr}, {Kim}, {Koo}, {Lee}, {Lee}, {Magakian}, {Movsessian}, {Nikogossian},
  {Pyo}, \& {Stanke}}]{fro11}
{Froebrich}, D., {Davis}, C.~J., {Ioannidis}, G., {et~al.} 2011,
  \href{http://dx.doi.org/10.1111/j.1365-2966.2010.18149.x}{\JournalTitle{\mnras},
  413, 480}

\bibitem[{{F{\"u}rst} {et~al.}(1989){F{\"u}rst}, {Hummel}, {Reich}, {Sofue},
  {Sieber}, {Reif}, \& {Dettmar}}]{fur89}
{F{\"u}rst}, E., {Hummel}, E., {Reich}, W., {et~al.} 1989, \JournalTitle{\aap},
  209, 361

\bibitem[{{F{\"u}rst} {et~al.}(1997){F{\"u}rst}, {Reich}, \&
  {Aschenbach}}]{fur97}
{F{\"u}rst}, E., {Reich}, W., \& {Aschenbach}, B. 1997, \JournalTitle{\aap},
  319, 655

\bibitem[{{F{\"u}rst} {et~al.}(1985){F{\"u}rst}, {Reich}, {Reich}, {Sofue}, \&
  {Handa}}]{fur85}
{F{\"u}rst}, E., {Reich}, W., {Reich}, P., {Sofue}, Y., \& {Handa}, T. 1985,
  \href{http://dx.doi.org/10.1038/314720a0}{\JournalTitle{\nat}, 314, 720}

\bibitem[{{Graham} {et~al.}(1991){Graham}, {Wright}, {Hester}, \&
  {Longmore}}]{graham91}
{Graham}, J.~R., {Wright}, G.~S., {Hester}, J.~J., \& {Longmore}, A.~J. 1991,
  \href{http://dx.doi.org/10.1086/115676}{\JournalTitle{\aj}, 101, 175}

\bibitem[{{Green} {et~al.}(1997){Green}, {Frail}, {Goss}, \&
  {Otrupcek}}]{gre97}
{Green}, A.~J., {Frail}, D.~A., {Goss}, W.~M., \& {Otrupcek}, R. 1997,
  \href{http://dx.doi.org/10.1086/118626}{\JournalTitle{\aj}, 114, 2058}

\bibitem[{{Green}(1989)}]{gre89}
{Green}, D.~A. 1989,
  \href{http://dx.doi.org/10.1093/mnras/238.3.737}{\JournalTitle{\mnras}, 238,
  737}

\bibitem[{{Green}(2019)}]{green19}
---. 2019,
  \href{http://dx.doi.org/10.1007/s12036-019-9601-6}{\JournalTitle{Journal of
  Astrophysics and Astronomy}, 40, 36}

\bibitem[{{Green} \& {Gull}(1982)}]{green82}
{Green}, D.~A., \& {Gull}, S.~F. 1982,
  \href{http://dx.doi.org/10.1038/299606a0}{\JournalTitle{\nat}, 299, 606}

\bibitem[{{Green} \& {Gull}(1989)}]{green89}
---. 1989,
  \href{http://dx.doi.org/10.1093/mnras/237.3.555}{\JournalTitle{\mnras}, 237,
  555}

\bibitem[{{Green} {et~al.}(1988){Green}, {Gull}, {Tan}, \& {Simon}}]{gre88}
{Green}, D.~A., {Gull}, S.~F., {Tan}, S.~M., \& {Simon}, A.~J.~B. 1988,
  \href{http://dx.doi.org/10.1093/mnras/231.3.735}{\JournalTitle{\mnras}, 231,
  735}

\bibitem[{{Gwinn} {et~al.}(1992){Gwinn}, {Moran}, \& {Reid}}]{gwi92}
{Gwinn}, C.~R., {Moran}, J.~M., \& {Reid}, M.~J. 1992,
  \href{http://dx.doi.org/10.1086/171493}{\JournalTitle{\apj}, 393, 149}

\bibitem[{{Hagen} {et~al.}(1955){Hagen}, {Lilley}, \& {McClain}}]{hag55}
{Hagen}, J.~P., {Lilley}, A.~E., \& {McClain}, E.~F. 1955,
  \href{http://dx.doi.org/10.1086/146096}{\JournalTitle{\apj}, 122, 361}

\bibitem[{{Harrus} \& {Slane}(1999)}]{har99}
{Harrus}, I.~M., \& {Slane}, P.~O. 1999,
  \href{http://dx.doi.org/10.1086/307138}{\JournalTitle{\apj}, 516, 811}

\bibitem[{{Harrus} {et~al.}(2004){Harrus}, {Slane}, {Hughes}, \&
  {Plucinsky}}]{har04}
{Harrus}, I.~M., {Slane}, P.~O., {Hughes}, J.~P., \& {Plucinsky}, P.~P. 2004,
  \href{http://dx.doi.org/10.1086/381355}{\JournalTitle{\apj}, 603, 152}

\bibitem[{{Hasegawa} {et~al.}(1987){Hasegawa}, {Gatley}, {Garden}, {Brand},
  {Ohishi}, {Hayashi}, \& {Kaifu}}]{has87}
{Hasegawa}, T., {Gatley}, I., {Garden}, R.~P., {et~al.} 1987,
  \href{http://dx.doi.org/10.1086/184941}{\JournalTitle{\apjl}, 318, L77}

\bibitem[{{Helfand} {et~al.}(2006){Helfand}, {Becker}, {White}, {Fallon}, \&
  {Tuttle}}]{hel06}
{Helfand}, D.~J., {Becker}, R.~H., {White}, R.~L., {Fallon}, A., \& {Tuttle},
  S. 2006, \href{http://dx.doi.org/10.1086/503253}{\JournalTitle{\aj}, 131,
  2525}

\bibitem[{{Hewitt} {et~al.}(2008){Hewitt}, {Yusef-Zadeh}, \& {Wardle}}]{hew08}
{Hewitt}, J.~W., {Yusef-Zadeh}, F., \& {Wardle}, M. 2008,
  \href{http://dx.doi.org/10.1086/588652}{\JournalTitle{\apj}, 683, 189}

\bibitem[{{Hollenbach} \& {Tielens}(1997)}]{hol97}
{Hollenbach}, D.~J., \& {Tielens}, A.~G.~G.~M. 1997,
  \href{http://dx.doi.org/10.1146/annurev.astro.35.1.179}{\JournalTitle{\araa},
  35, 179}

\bibitem[{{Hollenbach} \& {Tielens}(1999)}]{hol99}
---. 1999,
  \href{http://dx.doi.org/10.1103/RevModPhys.71.173}{\JournalTitle{Reviews of
  Modern Physics}, 71, 173}

\bibitem[{{Huang} \& {Thaddeus}(1986)}]{huang86}
{Huang}, Y.~L., \& {Thaddeus}, P. 1986,
  \href{http://dx.doi.org/10.1086/164649}{\JournalTitle{\apj}, 309, 804}

\bibitem[{{Jiang} {et~al.}(2010){Jiang}, {Chen}, {Wang}, {Su}, {Zhou},
  {Safi-Harb}, \& {DeLaney}}]{jia10}
{Jiang}, B., {Chen}, Y., {Wang}, J., {et~al.} 2010,
  \href{http://dx.doi.org/10.1088/0004-637X/712/2/1147}{\JournalTitle{\apj},
  712, 1147}

\bibitem[{{Junkes} {et~al.}(1992){Junkes}, {Fuerst}, \& {Reich}}]{jun92}
{Junkes}, N., {Fuerst}, E., \& {Reich}, W. 1992, \JournalTitle{\aaps}, 96, 1

\bibitem[{{Kaplan} {et~al.}(2017){Kaplan}, {Dinerstein}, {Oh}, {Mace}, {Kim},
  {Sokal}, {Pavel}, {Lee}, {Pak}, {Park}, {Sok Oh}, \& {Jaffe}}]{kap17}
{Kaplan}, K.~F., {Dinerstein}, H.~L., {Oh}, H., {et~al.} 2017,
  \href{http://dx.doi.org/10.3847/1538-4357/aa5b9f}{\JournalTitle{\apj}, 838,
  152}

\bibitem[{{Kazes}(1970)}]{kaz70}
{Kazes}, I. 1970, \JournalTitle{\aap}, 4, 111

\bibitem[{{Kilpatrick} {et~al.}(2016){Kilpatrick}, {Bieging}, \&
  {Rieke}}]{kil16}
{Kilpatrick}, C.~D., {Bieging}, J.~H., \& {Rieke}, G.~H. 2016,
  \href{http://dx.doi.org/10.3847/0004-637X/816/1/1}{\JournalTitle{\apj}, 816,
  1}

\bibitem[{{Koo} {et~al.}(2007){Koo}, {Moon}, {Lee}, {Lee}, \&
  {Matthews}}]{koo07}
{Koo}, B.-C., {Moon}, D.-S., {Lee}, H.-G., {Lee}, J.-J., \& {Matthews}, K.
  2007, \href{http://dx.doi.org/10.1086/510550}{\JournalTitle{\apj}, 657, 308}

\bibitem[{{Koo} {et~al.}(1993){Koo}, {Yun}, {Ho}, \& {Lee}}]{koo93}
{Koo}, B.-C., {Yun}, M.-S., {Ho}, P.~T.~P., \& {Lee}, Y. 1993,
  \href{http://dx.doi.org/10.1086/173303}{\JournalTitle{\apj}, 417, 196}

\bibitem[{{Koralesky} {et~al.}(1998){Koralesky}, {Frail}, {Goss}, {Claussen},
  \& {Green}}]{kor98}
{Koralesky}, B., {Frail}, D.~A., {Goss}, W.~M., {Claussen}, M.~J., \& {Green},
  A.~J. 1998, \href{http://dx.doi.org/10.1086/300508}{\JournalTitle{\aj}, 116,
  1323}

\bibitem[{{Kothes}(2013)}]{kothes13}
{Kothes}, R. 2013,
  \href{http://dx.doi.org/10.1051/0004-6361/201219839}{\JournalTitle{\aap},
  560, A18}

\bibitem[{{Kothes} {et~al.}(2003){Kothes}, {Reich}, {Foster}, \&
  {Byun}}]{kothes03}
{Kothes}, R., {Reich}, W., {Foster}, T., \& {Byun}, D.-Y. 2003,
  \href{http://dx.doi.org/10.1086/374219}{\JournalTitle{\apj}, 588, 852}

\bibitem[{{Lacey} {et~al.}(2001){Lacey}, {Lazio}, {Kassim}, {Duric}, {Briggs},
  \& {Dyer}}]{lacey01}
{Lacey}, C.~K., {Lazio}, T. J.~W., {Kassim}, N.~E., {et~al.} 2001,
  \href{http://dx.doi.org/10.1086/322372}{\JournalTitle{\apj}, 559, 954}

\bibitem[{{Le} {et~al.}(2017){Le}, {Pak}, {Kaplan}, {Mace}, {Lee}, {Pavel},
  {Jeong}, {Oh}, {Lee}, {Chun}, {Yuk}, {Pyo}, {Hwang}, {Kim}, {Park}, {Sok Oh},
  {Yu}, {Park}, {Minh}, \& {Jaffe}}]{le17}
{Le}, H. A.~N., {Pak}, S., {Kaplan}, K., {et~al.} 2017,
  \href{http://dx.doi.org/10.3847/1538-4357/aa6bf7}{\JournalTitle{\apj}, 841,
  13}

\bibitem[{{Leahy} {et~al.}(2014){Leahy}, {Green}, \& {Tian}}]{lea14}
{Leahy}, D., {Green}, K., \& {Tian}, W. 2014,
  \href{http://dx.doi.org/10.1093/mnras/stt2323}{\JournalTitle{\mnras}, 438,
  1813}

\bibitem[{{Leahy} \& {Tian}(2010)}]{lea10}
{Leahy}, D., \& {Tian}, W. 2010, in Astronomical Society of the Pacific
  Conference Series, Vol. 438, The Dynamic Interstellar Medium: A Celebration
  of the Canadian Galactic Plane Survey, ed. R.~{Kothes}, T.~L. {Landecker}, \&
  A.~G. {Willis}, 365

\bibitem[{{Lee} {et~al.}(2009){Lee}, {Moon}, {Koo}, {Lee}, \&
  {Matthews}}]{lee09}
{Lee}, H.-G., {Moon}, D.-S., {Koo}, B.-C., {Lee}, J.-J., \& {Matthews}, K.
  2009,
  \href{http://dx.doi.org/10.1088/0004-637X/691/2/1042}{\JournalTitle{\apj},
  691, 1042}

\bibitem[{{Lee} {et~al.}(2019){Lee}, {Koo}, {Lee}, {Burton}, \&
  {Ryder}}]{lee19}
{Lee}, Y.-H., {Koo}, B.-C., {Lee}, J.-J., {Burton}, M.~G., \& {Ryder}, S. 2019,
  \href{http://dx.doi.org/10.3847/1538-3881/ab0212}{\JournalTitle{\aj}, 157,
  123}

\bibitem[{{Lepp} \& {McCray}(1983)}]{lep83}
{Lepp}, S., \& {McCray}, R. 1983,
  \href{http://dx.doi.org/10.1086/161062}{\JournalTitle{\apj}, 269, 560}

\bibitem[{{Lockhart} \& {Goss}(1978)}]{loc78}
{Lockhart}, I.~A., \& {Goss}, W.~M. 1978, \JournalTitle{\aap}, 67, 355

\bibitem[{{Lockman}(1989)}]{loc89}
{Lockman}, F.~J. 1989,
  \href{http://dx.doi.org/10.1086/191383}{\JournalTitle{\apjs}, 71, 469}

\bibitem[{{Mazzalay} {et~al.}(2013){Mazzalay}, {Saglia}, {Erwin}, {Fabricius},
  {Rusli}, {Thomas}, {Bender}, {Opitsch}, {Nowak}, \& {Williams}}]{maz13}
{Mazzalay}, X., {Saglia}, R.~P., {Erwin}, P., {et~al.} 2013,
  \href{http://dx.doi.org/10.1093/mnras/sts204}{\JournalTitle{\mnras}, 428,
  2389}

\bibitem[{{Moffett} \& {Reynolds}(1994)}]{moffett94}
{Moffett}, D.~A., \& {Reynolds}, S.~P. 1994,
  \href{http://dx.doi.org/10.1086/175033}{\JournalTitle{\apj}, 437, 705}

\bibitem[{{Moorhouse} {et~al.}(1991){Moorhouse}, {Brand}, {Geballe}, \&
  {Burton}}]{moorhouse91}
{Moorhouse}, A., {Brand}, P.~W.~J.~L., {Geballe}, T.~R., \& {Burton}, M.~G.
  1991, \href{http://dx.doi.org/10.1093/mnras/253.4.662}{\JournalTitle{\mnras},
  253, 662}

\bibitem[{{Mouri}(1994)}]{mou94}
{Mouri}, H. 1994, \href{http://dx.doi.org/10.1086/174184}{\JournalTitle{\apj},
  427, 777}

\bibitem[{{Muller}(1959)}]{mul59}
{Muller}, C.~A. 1959, in IAU Symposium, Vol.~9, URSI Symp. 1: Paris Symposium
  on Radio Astronomy, ed. R.~N. {Bracewell}, 360

\bibitem[{{Oh} {et~al.}(2018){Oh}, {Pyo}, {Koo}, {Yuk}, {Kaplan}, {Lee},
  {Sokal}, {Mace}, {Park}, {Lee}, {Park}, {Hwang}, {Kim}, \& {Jaffe}}]{oh18}
{Oh}, H., {Pyo}, T.-S., {Koo}, B.-C., {et~al.} 2018,
  \href{http://dx.doi.org/10.3847/1538-4357/aabba4}{\JournalTitle{\apj}, 858,
  23}

\bibitem[{{Oliva} {et~al.}(1990){Oliva}, {Moorwood}, \& {Danziger}}]{oliva90}
{Oliva}, E., {Moorwood}, A.~F.~M., \& {Danziger}, I.~J. 1990,
  \JournalTitle{\aap}, 240, 453

\bibitem[{{Pak} {et~al.}(2004){Pak}, {Jaffe}, {Stacey}, {Bradford}, {Klumpe},
  \& {Keller}}]{pak04}
{Pak}, S., {Jaffe}, D.~T., {Stacey}, G.~J., {et~al.} 2004,
  \href{http://dx.doi.org/10.1086/421233}{\JournalTitle{\apj}, 609, 692}

\bibitem[{{Park} {et~al.}(2014){Park}, {Jaffe}, {Yuk}, {Chun}, {Pak}, {Kim},
  {Pavel}, {Lee}, {Oh}, {Jeong}, {Sim}, {Lee}, {Nguyen Le}, {Strubhar},
  {Gully-Santiago}, {Oh}, {Cha}, {Moon}, {Park}, {Brooks}, {Ko}, {Han}, {Nah},
  {Hill}, {Lee}, {Barnes}, {Yu}, {Kaplan}, {Mace}, {Kim}, {Lee}, {Hwang}, \&
  {Park}}]{par14}
{Park}, C., {Jaffe}, D.~T., {Yuk}, I.-S., {et~al.} 2014,
  \href{http://dx.doi.org/10.1117/12.2056431}{in \procspie, Vol. 9147,
  Ground-based and Airborne Instrumentation for Astronomy V}, 91471D

\bibitem[{{Parker} {et~al.}(2016){Parker}, {Boji{\v{c}}i{\'c}}, \&
  {Frew}}]{parker16}
{Parker}, Q.~A., {Boji{\v{c}}i{\'c}}, I.~S., \& {Frew}, D.~J. 2016,
  \href{http://dx.doi.org/10.1088/1742-6596/728/3/032008}{in Journal of Physics
  Conference Series, Vol. 728, Journal of Physics Conference Series}, 032008

\bibitem[{{Parker} {et~al.}(2005){Parker}, {Phillipps}, {Pierce}, {Hartley},
  {Hambly}, {Read}, {MacGillivray}, {Tritton}, {Cass}, {Cannon}, {Cohen},
  {Drew}, {Frew}, {Hopewell}, {Mader}, {Malin}, {Masheder}, {Morgan}, {Morris},
  {Russeil}, {Russell}, \& {Walker}}]{par05}
{Parker}, Q.~A., {Phillipps}, S., {Pierce}, M.~J., {et~al.} 2005,
  \href{http://dx.doi.org/10.1111/j.1365-2966.2005.09350.x}{\JournalTitle{\mnras},
  362, 689}

\bibitem[{{Paron} {et~al.}(2013){Paron}, {Weidmann}, {Ortega}, {Albacete
  Colombo}, \& {Pichel}}]{paron13}
{Paron}, S., {Weidmann}, W., {Ortega}, M.~E., {Albacete Colombo}, J.~F., \&
  {Pichel}, A. 2013,
  \href{http://dx.doi.org/10.1093/mnras/stt837}{\JournalTitle{\mnras}, 433,
  1619}

\bibitem[{{Pavlovic} {et~al.}(2014){Pavlovic}, {Dobardzic}, {Vukotic}, \&
  {Urosevic}}]{pav14}
{Pavlovic}, M.~Z., {Dobardzic}, A., {Vukotic}, B., \& {Urosevic}, D. 2014,
  \href{http://dx.doi.org/10.2298/SAJ1489025P}{\JournalTitle{Serbian
  Astronomical Journal}, 189, 25}

\bibitem[{{Persic} {et~al.}(1996){Persic}, {Salucci}, \& {Stel}}]{per96}
{Persic}, M., {Salucci}, P., \& {Stel}, F. 1996,
  \href{http://dx.doi.org/10.1093/mnras/281.1.27}{\JournalTitle{\mnras}, 281,
  27}

\bibitem[{{Radhakrishnan} {et~al.}(1972){Radhakrishnan}, {Goss}, {Murray}, \&
  {Brooks}}]{rad72}
{Radhakrishnan}, V., {Goss}, W.~M., {Murray}, J.~D., \& {Brooks}, J.~W. 1972,
  \href{http://dx.doi.org/10.1086/190249}{\JournalTitle{\apjs}, 24, 49}

\bibitem[{{Ranasinghe} \& {Leahy}(2017)}]{ran17}
{Ranasinghe}, S., \& {Leahy}, D.~A. 2017,
  \href{http://dx.doi.org/10.3847/1538-4357/aa7894}{\JournalTitle{\apj}, 843,
  119}

\bibitem[{{Ranasinghe} \& {Leahy}(2018{\natexlab{a}})}]{ran18b}
---. 2018{\natexlab{a}},
  \href{http://dx.doi.org/10.1093/mnras/sty817}{\JournalTitle{\mnras}, 477,
  2243}

\bibitem[{{Ranasinghe} \& {Leahy}(2018{\natexlab{b}})}]{ran18a}
---. 2018{\natexlab{b}},
  \href{http://dx.doi.org/10.3847/1538-3881/aab9be}{\JournalTitle{\aj}, 155,
  204}

\bibitem[{{Reid} {et~al.}(2014){Reid}, {Menten}, {Brunthaler}, {Zheng}, {Dame},
  {Xu}, {Wu}, {Zhang}, {Sanna}, {Sato}, {Hachisuka}, {Choi}, {Immer},
  {Moscadelli}, {Rygl}, \& {Bartkiewicz}}]{rei14}
{Reid}, M.~J., {Menten}, K.~M., {Brunthaler}, A., {et~al.} 2014,
  \href{http://dx.doi.org/10.1088/0004-637X/783/2/130}{\JournalTitle{\apj},
  783, 130}

\bibitem[{{Rieke}(2007)}]{rieke07}
{Rieke}, G.~H. 2007,
  \href{http://dx.doi.org/10.1146/annurev.astro.44.051905.092436}{\JournalTitle{\araa},
  45, 77}

\bibitem[{{Shan} {et~al.}(2018){Shan}, {Zhu}, {Tian}, {Zhang}, {Zhang}, {Wu},
  \& {Yang}}]{shan18}
{Shan}, S.~S., {Zhu}, H., {Tian}, W.~W., {et~al.} 2018,
  \href{http://dx.doi.org/10.3847/1538-4365/aae07a}{\JournalTitle{\apjs}, 238,
  35}

\bibitem[{{Sternberg} \& {Dalgarno}(1989)}]{ste89}
{Sternberg}, A., \& {Dalgarno}, A. 1989,
  \href{http://dx.doi.org/10.1086/167193}{\JournalTitle{\apj}, 338, 197}

\bibitem[{{Sternberg} \& {Neufeld}(1999)}]{ste99}
{Sternberg}, A., \& {Neufeld}, D.~A. 1999,
  \href{http://dx.doi.org/10.1086/307115}{\JournalTitle{\apj}, 516, 371}

\bibitem[{{Stil} {et~al.}(2006){Stil}, {Taylor}, {Dickey}, {Kavars}, {Martin},
  {Rothwell}, {Boothroyd}, {Lockman}, \& {McClure-Griffiths}}]{sti06}
{Stil}, J.~M., {Taylor}, A.~R., {Dickey}, J.~M., {et~al.} 2006,
  \href{http://dx.doi.org/10.1086/505940}{\JournalTitle{\aj}, 132, 1158}

\bibitem[{{Stupar} \& {Parker}(2011)}]{stu11}
{Stupar}, M., \& {Parker}, Q.~A. 2011,
  \href{http://dx.doi.org/10.1111/j.1365-2966.2011.18547.x}{\JournalTitle{\mnras},
  414, 2282}

\bibitem[{{Su} {et~al.}(2011){Su}, {Chen}, {Yang}, {Koo}, {Zhou}, {Lu},
  {Jeong}, \& {DeLaney}}]{su11}
{Su}, Y., {Chen}, Y., {Yang}, J., {et~al.} 2011,
  \href{http://dx.doi.org/10.1088/0004-637X/727/1/43}{\JournalTitle{\apj}, 727,
  43}

\bibitem[{{Tian} \& {Leahy}(2008)}]{tia08}
{Tian}, W.~W., \& {Leahy}, D.~A. 2008,
  \href{http://dx.doi.org/10.1111/j.1745-3933.2008.00557.x}{\JournalTitle{\mnras},
  391, L54}

\bibitem[{{Treffers}(1979)}]{treffers79}
{Treffers}, R.~R. 1979,
  \href{http://dx.doi.org/10.1086/183067}{\JournalTitle{\apjl}, 233, L17}

\bibitem[{{T{\"u}llmann} {et~al.}(2010){T{\"u}llmann}, {Plucinsky}, {Gaetz},
  {Slane}, {Hughes}, {Harrus}, \& {Pannuti}}]{tul10}
{T{\"u}llmann}, R., {Plucinsky}, P.~P., {Gaetz}, T.~J., {et~al.} 2010,
  \href{http://dx.doi.org/10.1088/0004-637X/720/1/848}{\JournalTitle{\apj},
  720, 848}

\bibitem[{{Umemoto} {et~al.}(2017){Umemoto}, {Minamidani}, {Kuno}, {Fujita},
  {Matsuo}, {Nishimura}, {Torii}, {Tosaki}, {Kohno}, {Kuriki}, {Tsuda},
  {Hirota}, {Ohashi}, {Yamagishi}, {Handa}, {Nakanishi}, {Omodaka}, {Koide},
  {Matsumoto}, {Onishi}, {Tokuda}, {Seta}, {Kobayashi}, {Tachihara}, {Sano},
  {Hattori}, {Onodera}, {Oasa}, {Kamegai}, {Tsuboi}, {Sofue}, {Higuchi},
  {Chibueze}, {Mizuno}, {Honma}, {Muller}, {Inoue}, {Morokuma-Matsui},
  {Shinnaga}, {Ozawa}, {Takahashi}, {Yoshiike}, {Costes}, \&
  {Kuwahara}}]{ume17}
{Umemoto}, T., {Minamidani}, T., {Kuno}, N., {et~al.} 2017,
  \href{http://dx.doi.org/10.1093/pasj/psx061}{\JournalTitle{\pasj}, 69, 78}

\bibitem[{{Whittet}(1992)}]{whittet92}
{Whittet}, D.~C.~B. 1992, {Dust in the galactic environment}

\bibitem[{{Williams} \& {Davies}(1954)}]{wil54}
{Williams}, D.~R.~W., \& {Davies}, R.~D. 1954,
  \href{http://dx.doi.org/10.1038/1731182a0}{\JournalTitle{\nat}, 173, 1182}

\bibitem[{{Wilson}(1970)}]{wil70}
{Wilson}, T.~L. 1970, \JournalTitle{\aplett}, 7, 95

\bibitem[{{Yuk} {et~al.}(2010){Yuk}, {Jaffe}, {Barnes}, {Chun}, {Park}, {Lee},
  {Lee}, {Wang}, {Park}, {Pak}, {Strubhar}, {Deen}, {Oh}, {Seo}, {Pyo}, {Park},
  {Lacy}, {Goertz}, {Rand}, \& {Gully-Santiago}}]{yuk10}
{Yuk}, I.-S., {Jaffe}, D.~T., {Barnes}, S., {et~al.} 2010,
  \href{http://dx.doi.org/10.1117/12.856864}{in \procspie, Vol. 7735,
  Ground-based and Airborne Instrumentation for Astronomy III}, 77351M

\bibitem[{{Yusef-Zadeh} {et~al.}(2003){Yusef-Zadeh}, {Wardle}, {Rho}, \&
  {Sakano}}]{yus03}
{Yusef-Zadeh}, F., {Wardle}, M., {Rho}, J., \& {Sakano}, M. 2003,
  \href{http://dx.doi.org/10.1086/345932}{\JournalTitle{\apj}, 585, 319}

\bibitem[{{Zhou} \& {Chen}(2011)}]{zho11}
{Zhou}, P., \& {Chen}, Y. 2011,
  \href{http://dx.doi.org/10.1088/0004-637X/743/1/4}{\JournalTitle{\apj}, 743,
  4}

\bibitem[{{Zhou} {et~al.}(2009){Zhou}, {Chen}, {Su}, \& {Yang}}]{zho09}
{Zhou}, X., {Chen}, Y., {Su}, Y., \& {Yang}, J. 2009,
  \href{http://dx.doi.org/10.1088/0004-637X/691/1/516}{\JournalTitle{\apj},
  691, 516}

\bibitem[{{Zhu} {et~al.}(2014){Zhu}, {Tian}, \& {Zuo}}]{zhu14}
{Zhu}, H., {Tian}, W.~W., \& {Zuo}, P. 2014,
  \href{http://dx.doi.org/10.1088/0004-637X/793/2/95}{\JournalTitle{\apj}, 793,
  95}

\end{thebibliography}


\clearpage
\startlongtable
\begin{deluxetable}{cccccccc}
\tabletypesize{\scriptsize}
\tablewidth{0pt}
\tablecolumns{8}
\tablecaption{Properties of \hh\ Lines \label{tab-prop}}
\tablehead{
\multirow{2}{*}{SNR}	& \multirow{2}{*}{Slit}	&   \colhead{Slit Position} &
\colhead{$v_{\rm LSR}$}  &
\colhead{FWHM}	&
\multirow{2}{*}{$\dfrac{2-1~S(1)}{1-0~S(1)}$} &
\multirow{2}{*}{$\dfrac{1-0~S(2)}{1-0~S(0)}$} &  
\multirow{2}{*}{$\dfrac{1-0~S(0)}{1-0~S(1)}$} \\
    & 	&
\colhead{[ $\alpha$(J2000) ~~~~ $\delta$(J2000) ]}  &
\colhead{($\kms$)}  &
\colhead{($\kms$)}  &   &   &   \\
\colhead{(1)}	&   \colhead{(2)}	&   \colhead{(3)}	&   \colhead{(4)}	&
\colhead{(5)}	&   \colhead{(6)}	&   \colhead{(7)}	&   \colhead{(8)}
}
\startdata
G9.9$-$0.8
    &	NW	&   18:10:26.33 ~ $-$20:39:41.9 
    &	$+$30.4 (0.1)	&	4.4 (0.1)   &	0.31 (0.04) &   1.62 (0.19)  &   0.49 (0.04)  \\
G11.2$-$0.3
    &	N	&   18:11:26.64 ~ $-$19:21:52.2 
    &	$+$47.3 (0.2)	&	12.6 (0.4)  &   $<$0.16   	&   $<$15.2 	&   0.18 (0.04) \\
    &	S	&   18:11:28.85 ~ $-$19:27:51.5 
    &	$+$46.5 (0.1)	&	13.2 (0.3)	&   0.11 (0.03)	&   $<$5.54 	&   0.15 (0.02) \\
    &	SE	&   18:11:32.35 ~ $-$19:27:12.0 
    &	$+$49.2 (0.1)	&	12.0 (0.1)	&	0.12 (0.01)	&   2.30 (0.07) &   0.17 (0.01) \\
    &	NE	&   18:11:34.82 ~ $-$19:23:43.0 
    &	$+$48.9 (0.1)	&	11.4 (0.1)	&	0.07 (0.02) &   2.84 (0.47)	&   0.16 (0.01)  \\
G13.5$+$0.2	
    &	S	&   18:14:20.85 ~ $-$17:13:57.4 
    &	$+$40.1 (0.9)	&	32.8 (2.1)	&	$<$0.28     &   1.67 (1.02) &   0.13 (0.05)  \\
G16.0$-$0.5	
    &	E1	&   18:22:13.88 ~ $-$15:18:15.9 
    &	$+$50.0 (0.3)	&	17.5 (0.8)	&	$<$0.18     &   2.30 (0.94) &   0.20 (0.06) \\
	&	E2	&   18:22:16.47 ~ $-$15:16:27.5 
	&	$+$52.7 (0.4)	&	17.1 (0.8)	&	$<$0.11     &   1.61 (0.54) &   0.19 (0.05)  \\
G18.1$-$0.1	
    &	NW	&   18:24:29.08 ~ $-$13:10:20.7 
    &	$+$73.1 (0.4)	&	26.6 (0.8)	&	0.09 (0.05)	&   1.60 (0.40) &   0.19 (0.04) \\
	&	SE	&   18:24:49.55 ~ $-$13:12:25.9 
	&	$+$85.3 (0.2)	&	13.5 (0.4)	&	$<$0.09		&   1.99 (0.53) &   0.20 (0.04)  \\
G18.9$-$1.1   
    &	C (p)\tablenotemark{a}  &   18:29:31.05 ~ $-$12:51:23.7 
    &	$+$74.0 (1.1)	&	26.0 (2.2)	&	0.13 (0.05) &   2.26 (0.78)	&   0.18 (0.06) \\
    &	C (s)\tablenotemark{a}  &   -
    &	$+$53.3 (1.1)	&	12.0 (2.1)	&   -           &   -           &   -           \\
	&	NE	&   18:30:04.44 ~ $-$12:51:43.0 
	&	$+$69.7 (0.5)	&	40.7 (1.3)	&	0.15 (0.06) &   2.71 (0.80)	&   0.16 (0.04)  \\
Kes 69		
    &	SE1	&   18:33:01.92 ~ $-$10:13:43.6 
    &	$+$59.0 (0.4)	&	40.7 (1.0)	&   0.16 (0.03) &	1.70 (0.39) &   0.19 (0.03) \\
	&	SE2	&   18:33:14.96 ~ $-$10:12:12.8 
	&	$+$61.7 (0.3)	&	28.5 (0.7)	&   0.15 (0.03) &	1.62 (0.30) &   0.23 (0.04) \\
    &	NE (p)\tablenotemark{a}	&   18:33:10.12 ~ $-$10:00:52.0 
    &	$+$77.0 (0.8)	&	28.8 (1.6)	&   0.12 (0.05) &	2.51 (0.91) &   0.18 (0.06) \\
    &   NE (s)\tablenotemark{a} &   -
    &	$+$33.4 (6.6)	&	43.9 (15.2)	&   -           &   -           &   -           \\
Kes 73	
    &	W	&   18:41:12.15 ~ $-$04:56:42.6 
    &	$+$99.4 (0.1)	&	10.9 (0.2)	&	0.09 (0.02)	&   2.24 (0.34) &   0.18 (0.02) \\
3C 391		
    &	NW	&   18:49:16.00 ~ $-$00:55:04.4 
    &	$+$99.7 (0.8)	&	53.3 (2.1)	&	0.11 (0.04) &   2.66 (1.00)	&   0.13 (0.04) \\
    &	NE (p)\tablenotemark{a}  &   18:49:27.86 ~ $-$00:55:01.4 
    &	$+$96.4 (0.4)	&	29.2 (1.8)	&	0.09 (0.02) &   2.13 (0.29)	&   0.19 (0.04) \\
    &	NE (s)\tablenotemark{a}  &   -
    &	$+$76.2 (4.7)	&	50.6 (4.0)	&   -           &   -           &   -           \\
    &	SW (p)\tablenotemark{a}  &   18:49:23.26 ~ $-$00:57:41.0 
    &	$+$107.0 (0.1)	&	25.4 (0.4)	&	0.09 (0.01) &   1.96 (0.11)	&   0.20 (0.01) \\
    &	SW (s)\tablenotemark{a}  &   -
    &	$+$84.0 (1.0)	&	58.0 (1.1)	&   -           &   -           &   -           \\
G32.1$-$0.9	
    &	NW	&   18:52:40.73 ~ $-$00:56:57.4 
    &	$+$84.5 (0.2)	&	18.5 (0.5)	&	$<$0.10     &   1.74 (0.45) &   0.22 (0.05) \\
	&	C	&   18:53:42.22 ~ $-$01:02:01.8 
	&	$+$100.4 (0.1)	&	10.7 (0.2)	&	$<$0.07     &   1.95 (0.32) &   0.21 (0.03)  \\
Kes 78		
    &	E	&   18:51:46.34 ~ $-$00:11:27.7 
    &	$+$89.5 (0.2)	&	22.4 (0.4)	&	0.11 (0.03) &   2.10 (0.34)	&   0.19 (0.03)  \\
G33.2$-$0.6	
    &	W	&   18:53:25.70 ~ $+$00:01:09.4 
    &	$+$81.7 (0.5)	&	26.4 (1.2)	&	$<$0.17     &   2.26 (0.79) &   0.20 (0.06)  \\
W44			
    &	W	&   18:55:17.10 ~ $+$01:21:50.0 
    &	$+$46.2 (0.1)	&	38.4 (0.1)	&	0.10 (0.01)	&   1.84 (0.04) &   0.21 (0.01) \\
	&	N	&   18:55:39.80 ~ $+$01:37:56.4 
	&	$+$41.2 (0.1)	&	17.3 (0.3)	&	0.14 (0.02)	&   2.04 (0.25) &   0.24 (0.02) \\
	&	C	&   18:56:14.70 ~ $+$01:20:52.7 
	&	$+$26.7 (0.2)	&	20.0 (0.4)	&	0.06 (0.04)	&   1.61 (0.32) &   0.24 (0.03) \\
	&	S	&   18:56:35.46 ~ $+$01:06:44.1 
	&	$+$48.9 (0.1)	&	24.7 (0.3)	&	0.16 (0.02)	&   1.82 (0.25) &   0.20 (0.02) \\
3C 396		
    &	W1	&   19:03:55.86 ~ $+$05:25:36.9 
    &	$+$55.9 (0.1)	&	10.6 (0.2)	&	0.08 (0.02)	&   $<$5.83 	&   0.23 (0.02) \\
	&	W2	&   19:03:56.94 ~ $+$05:24:38.3 
	&	$+$57.5 (0.3)	&	11.9 (0.5)	&	$<$0.10 	&   $<$13.9 	&   0.13 (0.04)  \\
W49B		
    &	E1	&   19:11:16.14 ~ $+$09:05:04.9 
    &	$+$65.5 (0.1)	&	8.7 (0.1)	&	0.07 (0.01)	&   2.44 (0.24) &   0.18 (0.01) \\
	&	E2	&   19:11:16.91 ~ $+$09:06:17.5 
	&	$+$64.3 (0.1)	&	9.2 (0.1)	&	0.10 (0.01)	&   2.19 (0.17) &   0.18 (0.01) \\
	&	W	&   19:11:00.15 ~ $+$09:05:07.9 
	&	$+$61.7 (0.1)	&	25.7 (0.2)	&	0.11 (0.01)	&	2.41 (0.16) &   0.20 (0.01) \\
	&	N	&   19:11:06.46 ~ $+$09:06:47.9 
	&	$+$64.1 (0.1)	&	9.3 (0.3)	&	0.08 (0.03)	&   1.95 (0.49) &   0.18 (0.04) \\
HC 40		
    &	NW	&   19:32:14.18 ~ $+$19:07:31.4 
    &	$+$44.2 (0.4)	&	23.6 (0.9)	&	0.12 (0.05)	&   $<$3.08 	&   0.19 (0.05) \\
\enddata
\tablecomments{
Column (1): SNR name.
Column (2): slit name.
Column (3): central coordinate of the slit.
Column (4): central velocity of the \hh\ 1--0 S(1) 2.122~\micron\ line
in the LSR frame.
The uncertainties in parentheses are formal 1$\sigma$ statistical errors. 
The uncertainty in absolute wavelength calibration is about $1~\kms$.
Column (5): FWHM of the \hh\ 1--0 S(1) lines corrected 
for the instrument broadening (7 $\kms$).
Columns (6)--(8): extinction-corrected flux ratios (see text).
The numbers in parentheses are $1\sigma$ errors, 
whereas the numbers with ``$<$'' are $3\sigma$ upper limits.
}
\tablenotetext{a}{
The \hh\ line profiles are fitted by two Gaussian components:   
primary (p) and secondary (s).
See Figure~\ref{fig-spec} and the explanation in Section~\ref{sec-obs-nir}.
}
\end{deluxetable}

\clearpage
\startlongtable
\begin{deluxetable}{c|ccccc|cc}
\tabletypesize{\scriptsize}
\tablewidth{0pt}
\tablecolumns{8}
\tablecaption{Kinematic Distances of 16 Galactic SNRs \label{tab-dis}}
\tablehead{
\multirow{3}{*}{SNR}	& \multicolumn{5}{c}{Literature}	&
\multicolumn{2}{c}{\hh}   \\
\cline{2-6} \cline{7-8}
    	&
\colhead{$v_{\rm LSR}$}  &	\multirow{2}{*}{KDA}	&   \colhead{$d$}	&
\multirow{2}{*}{Method}    &	\multirow{2}{*}{Reference} &
\colhead{$v_{\rm LSR}$}   &   \colhead{$d$} \\
    	&
\colhead{(km s$^{-1}$)}	&       	&   \colhead{(kpc)}	&
        &       	&
\colhead{(km s$^{-1}$)}	&   \colhead{(kpc)}   \\
\colhead{(1)}	&   \colhead{(2)}	&   \colhead{(3)}	&   \colhead{(4)}	&
\colhead{(5)}	&   \colhead{(6)}	&   \colhead{(7)}	&   \colhead{(8)}
}
\startdata
G9.9$-$0.8    &   $+31$   &   N?   &   3.8 &   CO, $\Sigma$-$D$  &   1   &   
                $+30~(\pm 1)$\tablenotemark{a}   &   $3.8~(\pm 0.1)$\tablenotemark{a} \\
G11.2$-$0.3   &   $+45$   &   N   &   4.6 &   \hi\  &   2    &   
                $+48~(\pm 2)$   &   $4.7~(\pm 0.1)$  \\
G13.5$+$0.2   &   ...         &   F?   &   $13\pm7$    &   $\Sigma$-$D$  &   3    &   
                $+40~(\pm 1)$   &   $12.4~(\pm 0.1)$  \\
G16.0$-$0.5   &   ...         &   N?   &   $8\pm4$    &   $\Sigma$-$D$  &   3    &   
                $+51~(\pm 2)$   &   $4.1~(\pm 0.1)$ \\
G18.1$-$0.1   &   $+53$, $+100$  &   N   &   4.0, 6.2    &   \hi, CO  &   4, 5, 6    &   
                $+73$--$+85$   &   5.0--5.5 \\
G18.9$-$1.1   &   $+18$   &   N   &   1.7--1.8\tablenotemark{b}    &   \hi, RCS  &   7, 8    &   
                $+70~(\pm 1)$   &   $4.7~(\pm 0.1)$   \\
Kes 69     &   $+85$   &   N   &   5.2    &   \hi, CO, OH  &   9, 10, 11    &   
                $+61~(\pm 2)$   &   $4.1~(\pm 0.1)$ \\
Kes 73      &   $+100$  &   N   &   5.9    &   \hi, CO  &   1, 6    &   
                $+99~(\pm 1)$   &   $5.8~(\pm 0.1)$ \\
3C 391      &   $+100$--$+110$  &   TP   &   7.1    &   \hi, CO, OH  &   1, 12, 13    &   
                $+100~(\pm 1)$  &   $7.1~(\pm 0.1)$ \\
G32.1$-$0.9   &   ...         &   N?   &   4.6    &   Sedov  &   14    &   
                $+85~(\pm 1)$   &   $5.0~(\pm 0.1)$ \\
Kes 78      &   $+82$--$+86$   &   N   &   4.9--5.1    &   \hi, CO, OH  &   6, 15, 16    & 
                $+90~(\pm 1)$   &   $5.4~(\pm 0.1)$ \\
G33.2$-$0.6   &   ...         &   N?   &   $6\pm3$    &   $\Sigma$-$D$  &   3    &
                $+82~(\pm 1)$   &   $4.9~(\pm 0.1)$ \\
W44         &   $+42$--$+50$   &   N   &   2.6--3.1    &   \hi, OH  &   6, 17, 18    & 
                $+41$--$+49$   &   2.6--3.0 \\
3C 396      &   $+69$, $+84$   &   F   &   8.6, 6.8    &   \hi, CO  &   6, 19, 20    & 
                $+56~(\pm 2)$   &   $9.5~(\pm 0.1)$ \\
W49B        &   $+13$, $+40$   &   F   &   11.3, 9.6    &   \hi, CO  &   1, 6, 21    & 
                $+64~(\pm 2)$   &   $7.5~(\pm 0.2)$ \\
HC 40       &   $+36$--$+44$   &   F   &   5.4--6.8    &   \hi, CO  &   13, 22    & 
                $+44~(\pm 1)$   &   $5.4~(\pm 0.1)$ \\
\enddata
\tablecomments{
Column (1): SNR name.
Column (2): systemic velocity of SNR from the literature.
Column (3): kinematic distance ambiguity,
indicating whether the SNR is at the near (N), far (F),
or tangent point (TP) distance.
Those with ``?'' indicate that
the distance ambiguity had not been resolved 
in previous \hi\ absorption studies
and that we have adopted either near or far distance 
based on other studies (see text).
Column (4): kinematic distance derived from 
the systemic velocity in Column (2). For those without 
the velocity information, we list the distances 
from other methods listed in Column (5). 
Column (5): distance estimation method:
\hi\ absorption (\hi), CO emission (CO),
$\Sigma$-$D$ relation ($\Sigma$-$D$),
Sedov analysis (Sedov), and
red clump star (RCS) method (see text).
Column (6): references for the systemic velocity 
or the distance. 
Column (7): systemic velocity of the SNR
derived from the central velocities of the \hh\ lines.
The numbers in the parentheses are $1\sigma$ uncertainties, which are 
obtained by adding the absolute wavelength calibration uncertainty ($1~\kms$)
and the standard deviation of the central velocities in quadrature.
For the SNRs with multiple slit observations
and with velocity spread larger than 5 \kms, 
the minimum and maximum central velocities are given
(see Section~\ref{sec-res-dis}).
Column (8): kinematic distance derived from the systemic velocity in Column (7).
}
\tablenotetext{a}{
The \hh\ emission is most likely arising from a PDR,
and its association with the SNR is not clear
(see Section~\ref{sec-dis-g9.9}).
}
\tablenotetext{b}{
1.7 kpc is from \hi\ absorption, and 1.8 kpc is from the RCS method.
}
\tablerefs{
(1) \citealt{kil16};
(2) \citealt{gre88};
(3) \citealt{pav14};
(4) \citealt{paron13};
(5) \citealt{lea14};
(6) \citealt{ran18a};
(7) \citealt{fur89};
(8) \citealt{shan18};
(9) \citealt{hew08};
(10) \citealt{tia08};
(11) \citealt{zho09};
(12) \citealt{fra96};
(13) \citealt{ran17};
(14) \citealt{fol97};
(15) \citealt{kor98};
(16) \citealt{zho11};
(17) \citealt{cas75};
(18) \citealt{cla97};
(19) \citealt{lee09};
(20) \citealt{su11};
(21) \citealt{zhu14};
(22) \citealt{jun92}.
}
\end{deluxetable}


\clearpage
\begin{figure}
\includegraphics[width=\textwidth]{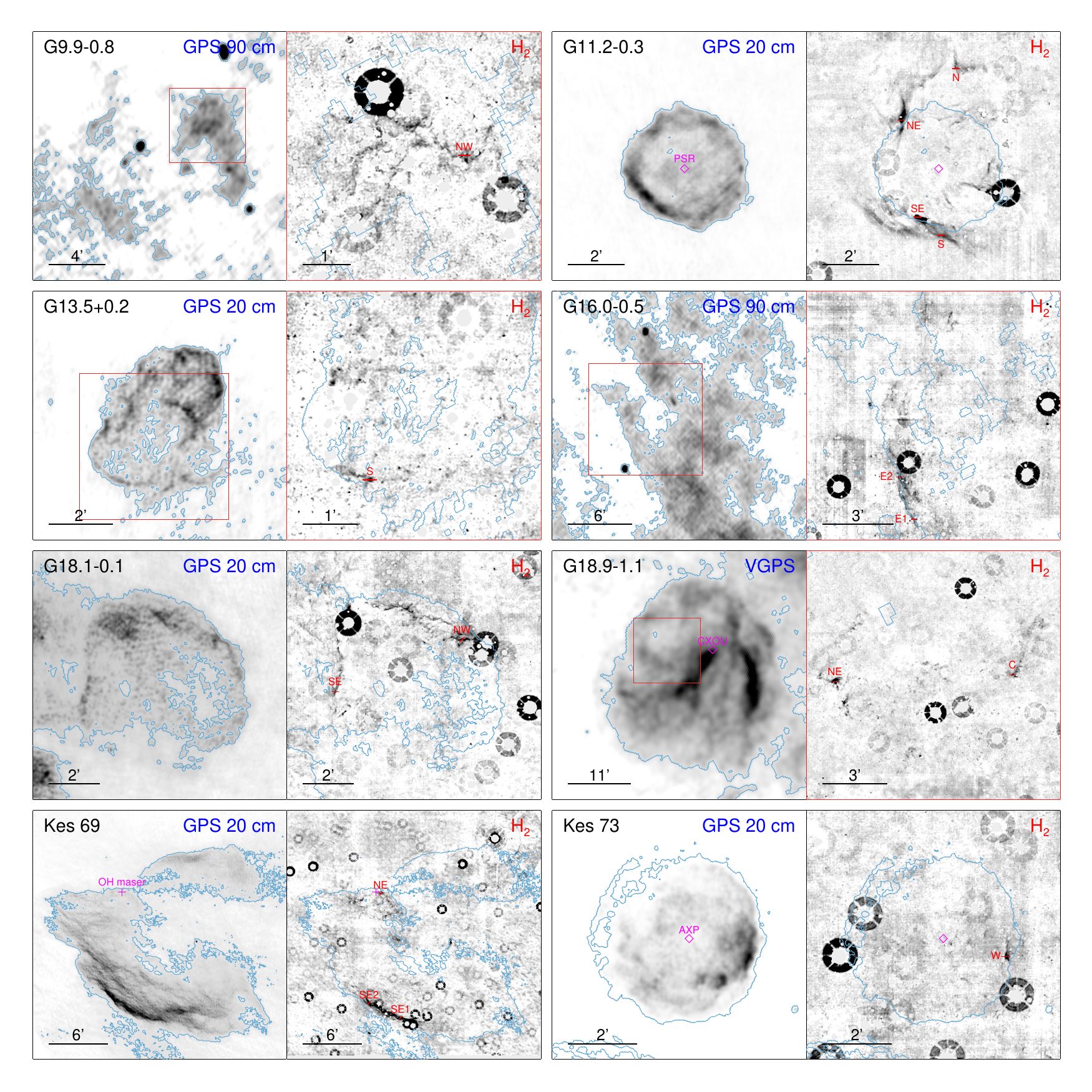}
\caption{
    Radio and \hh\ $2.122~\micron$ narrowband images of the 16 SNRs
    studied in this paper.
    North is up, and east is to the left.
    The radio images are from either
    the VLA 20/90~cm Galactic Plane Survey \citep[GPS 20/90cm;][]{hel06} or
    the VLA 21~cm Galactic Plane Survey \citep[VGPS;][]{sti06},
    while the \hh\ images are from \citet{lee19}.
    The gray scales are linear,
    and the blue contours represent the SNR boundary in radio continuum.
    In the radio images,
    the red box marks the boundary of the \hh\ image of the source and 
    the magenta symbols represent the locations of X-ray/radio sources
    associated with the SNR: 
    crosses = OH masers; open diamonds = pulsars or point-like X-ray sources.   
    The slit positions for the NIR spectroscopy 
    are marked on the \hh\ images (red bars).
    The slit names represent
    the directions from the SNR's center to the slits
    (i.e., N, S, E, and W indicate north, south, east, and west, respectively,
    whereas C indicates the central area inside the SNR boundary).
} \label{fig-slitpos-1}
\end{figure}

\addtocounter{figure}{-1}
\clearpage
\begin{figure}
\includegraphics[width=\textwidth]{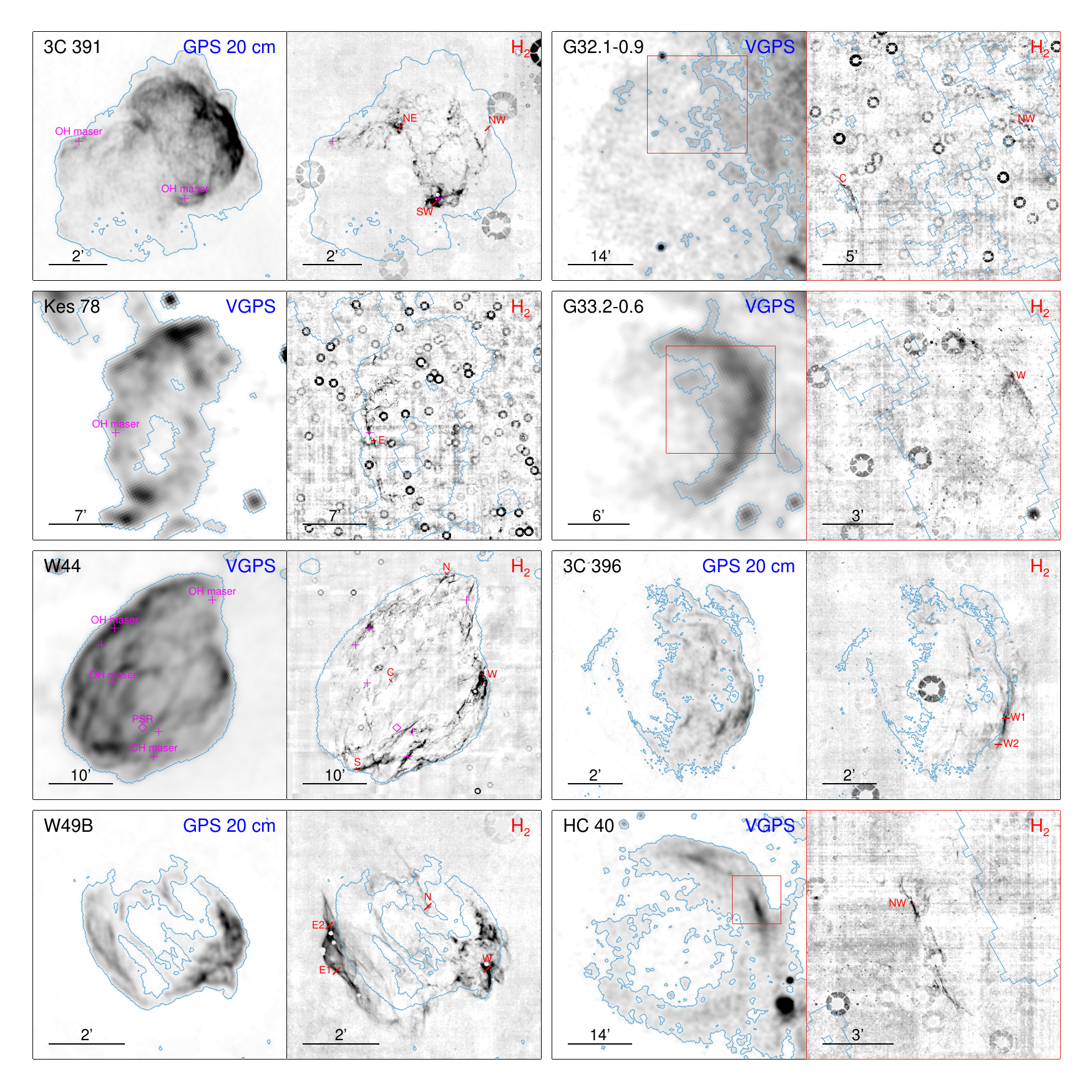}
\caption{
    (Continued)
} \label{fig-slitpos-2}
\end{figure}

\clearpage
\begin{figure}
\includegraphics[width=\textwidth]{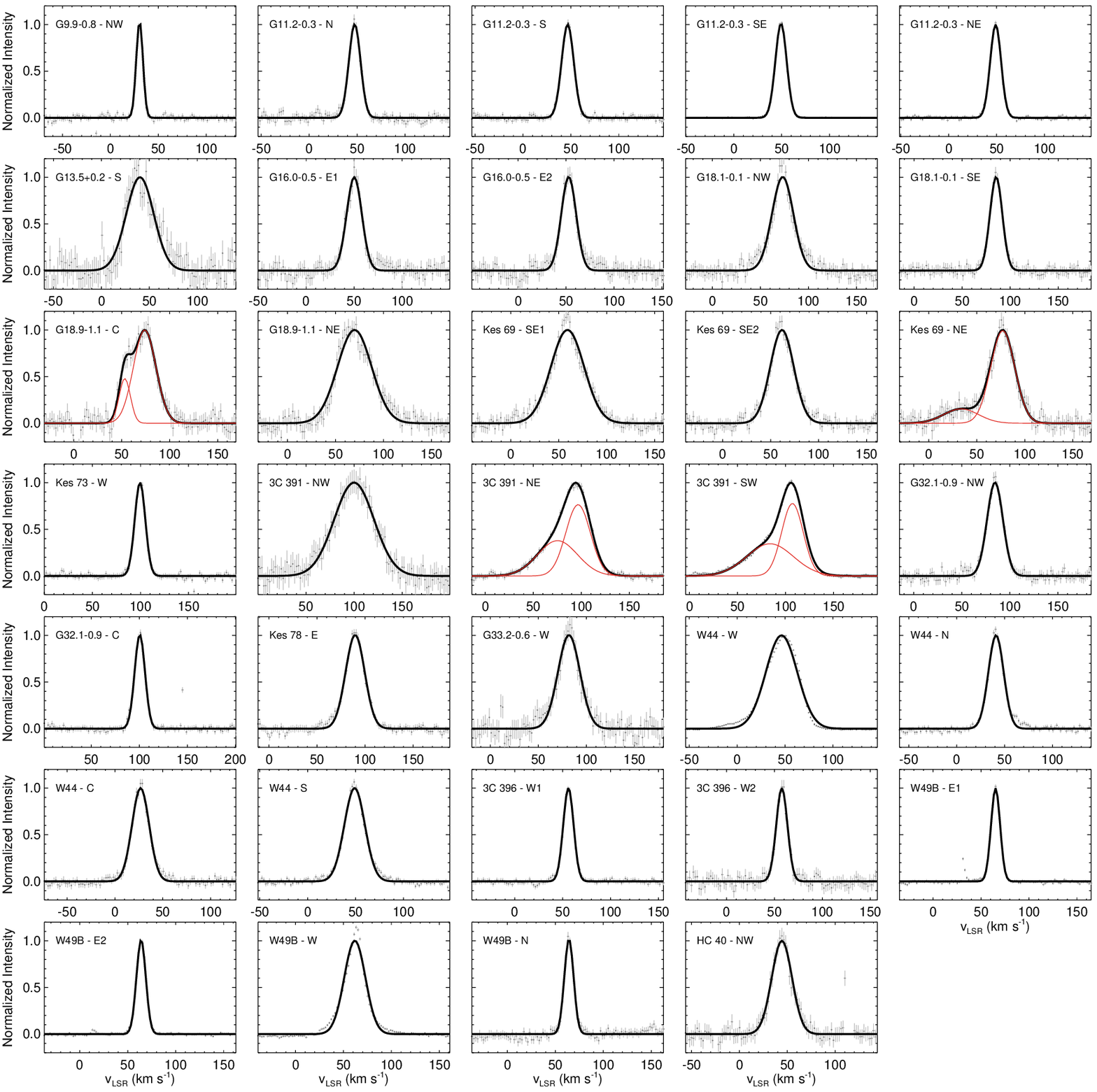}
\caption{
    \hh\ 1--0 S(1) 2.122~\micron\ spectra of the 34 targets
    in Table~\ref{tab-prop}.
    The gray dots with vertical bars represent the observed spectra
    with $1\sigma$ uncertainties,
    while the black solid lines are the model profiles
    resulting from Gaussian fittings.
    For four targets, G18.9$-$1.1-C, Kes 69-NE, 3C 391-NE, and 3C 391-SW,
    the spectra are fitted by two Gaussian components
    (see Section~\ref{sec-obs-nir}),
    and each component is shown by a red solid line.
} \label{fig-spec}
\end{figure}

\clearpage
\begin{figure}
\includegraphics[width=\textwidth]{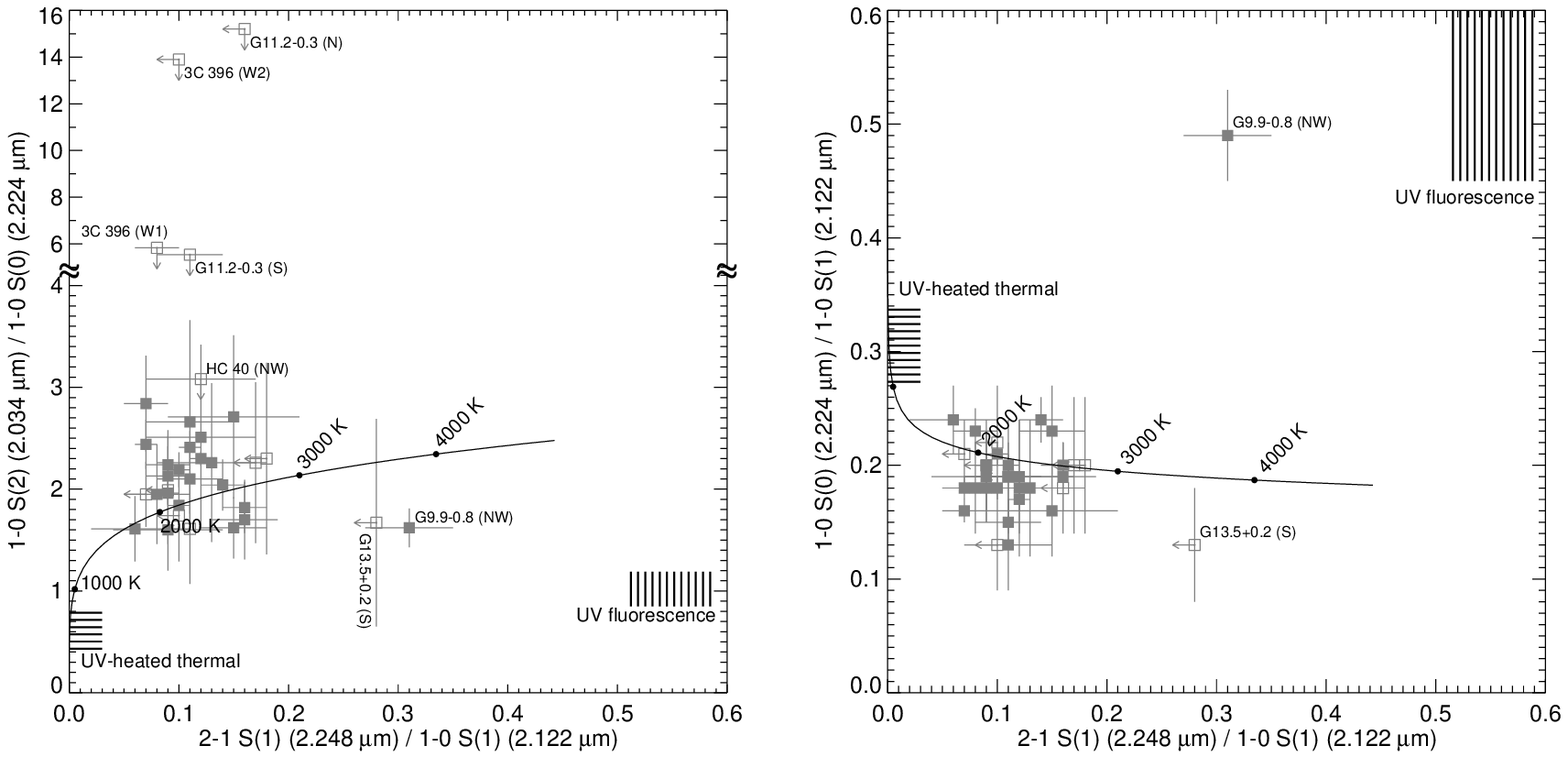}
\caption{
    Flux ratios of \hh\ emission lines:
    2--1 S(1) (2.248~\micron) / 1--0 S(1) (2.122~\micron)
	vs. 1--0 S(2) (2.034~\micron) / 1--0 S(0) (2.224~\micron) (left), and
	2--1 S(1) (2.248~\micron) / 1--0 S(1) (2.122~\micron)
	vs. 1--0 S(0) (2.224~\micron) / 1--0 S(1) (2.122~\micron) (right).
    The filled squares represent the extinction-corrected flux ratios of
    34 targets (Table~\ref{tab-prop}),
    while the open squares with arrows indicate
	the $3\sigma$ upper limits of the ratios.
	The black solid line represents the flux ratios of thermal gas in LTE
	at $T=1000$--4000 K.
	The vertical and horizontal stripes indicate the areas of the diagram
	corresponding to nonthermal UV fluorescence \citep{bla87} and
	UV-heated thermal excitation models \citep{ste89}, respectively.
	Note that the $y$-axis of the left panel has a discontinuity at $\sim 5$.
} \label{fig-ratio}
\end{figure}

\clearpage
\begin{figure}
\includegraphics[width=\textwidth]{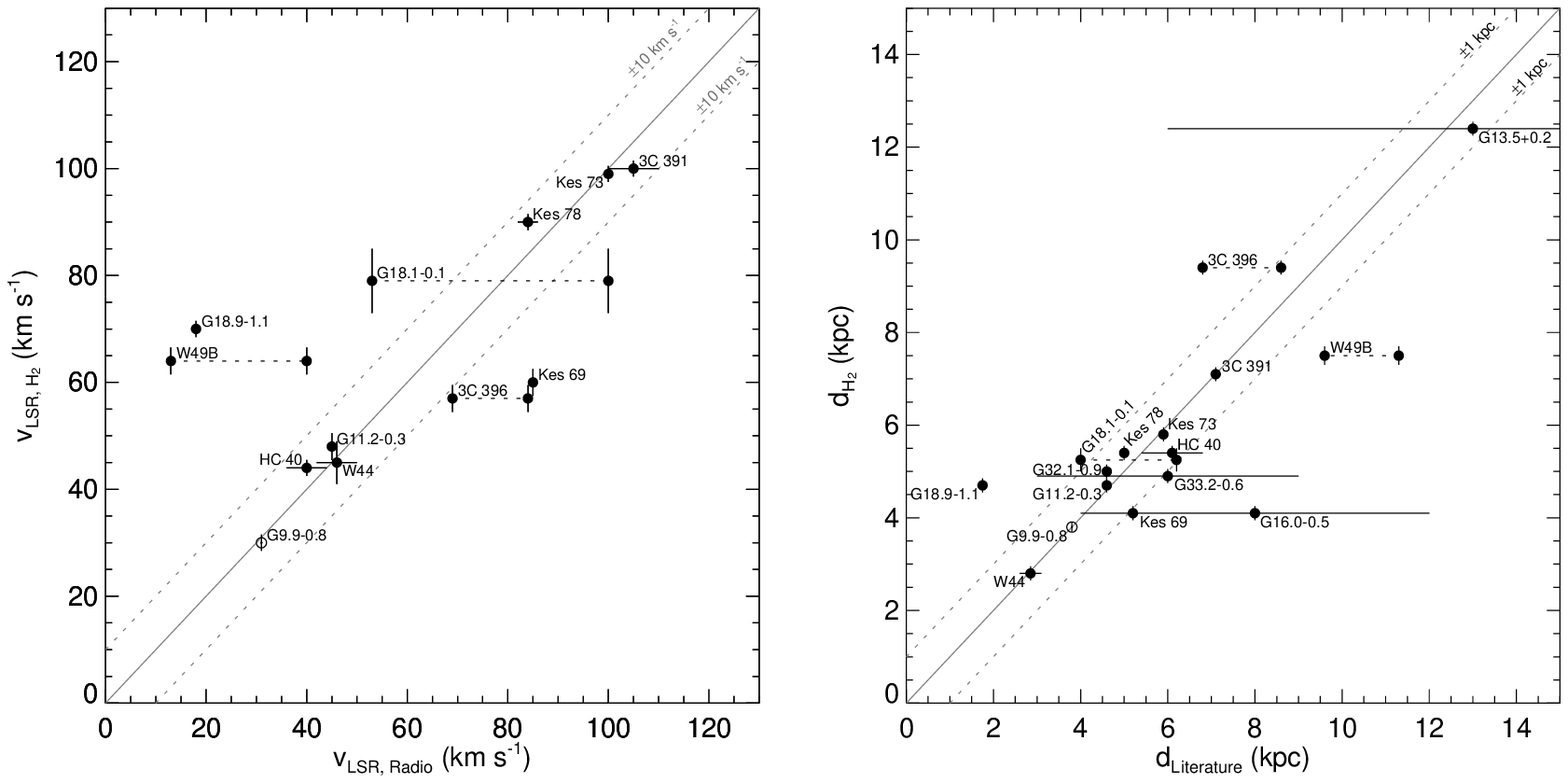}
\caption{
    Left: comparisons of the systemic velocities of the 16 SNRs  
    obtained from \hh\ 2.122~\micron\ lines ($v_{\rm LSR, H_2}$) 
    with the previous results ($v_{\rm LSR, radio}$). 
    The gray solid line represents the perfect agreement,
    whereas the dashed lines indicate $\pm 10~\kms$ differences between them.
    For the sources with very different suggested systemic velocities    
    in previous studies (G18.1$-$0.1, 3C 396, W49B), 
    we mark their minimum and maximum
    $v_{\rm LSR, radio}$ connected by horizontal dashed lines.
    G9.9$-$0.8 is marked with an open symbol because 
    the association between the \hh\ emission and the SNR is suspicious
    (see Section~\ref{sec-dis-g9.9}).
    Right: same as the left panel, but for the distances. 
    The dashed lines indicate $\pm 1$~kpc differences
    from the perfect agreement.
} \label{fig-vel}
\end{figure}

\clearpage
\begin{figure}
\includegraphics[width=0.5\textwidth]{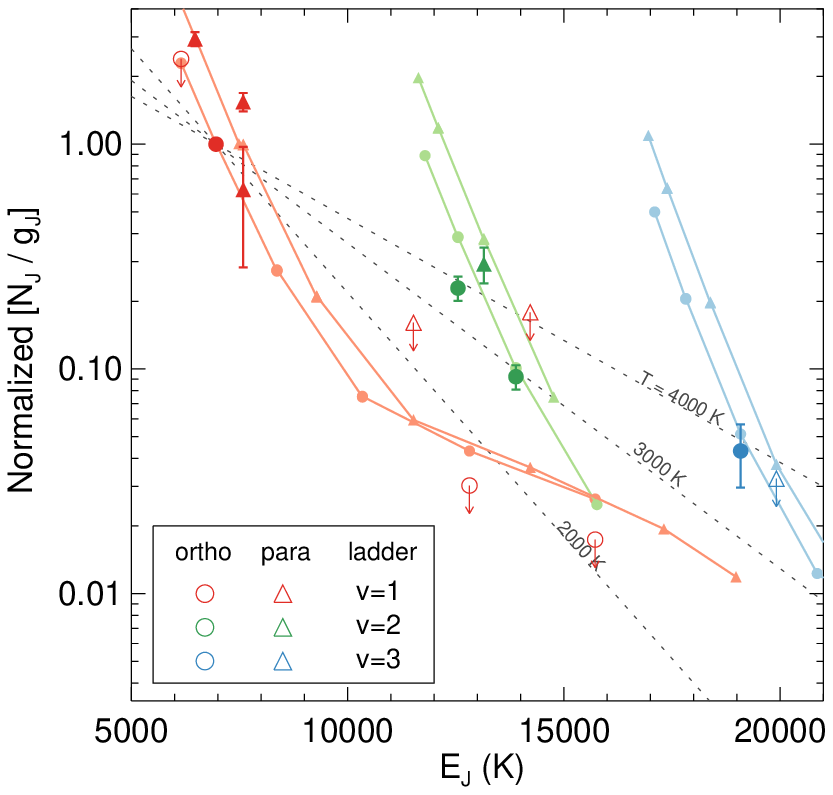}
\caption{
    Excitation diagram of \hh\ lines detected in G9.9$-$0.8.
	The {\it y}-axis is the column density ($N_{v,J}$)
	divided by the level degeneracy ($g_{J}$),
	and is normalized by that of the \hh\ 1--0 S(1) line.
	The symbols with a downward-pointing arrow denote $3\sigma$ upper limits. 
	The dashed lines represent the Boltzmann distribution with
	$T=2000$, 3000, and 4000 K.
	The solid lines represent the level populations expected from
	a warm, diffuse PDR
	with a significant contribution from collisional excitation 
	(i.e., the ``bh3d'' model of \citealt{dra96}
	with $n_{\rm H}=10^{2}$ cm$^{-3}$, $\chi=10$, and $T_0=1000$ K,
	where $\chi$ is a parameter characterizing the UV intensity and
	$T_0$ is the temperature at the PDR boundary).
} \label{fig-g99_lev}
\end{figure}

\clearpage
\begin{figure}
\includegraphics[width=\textwidth]{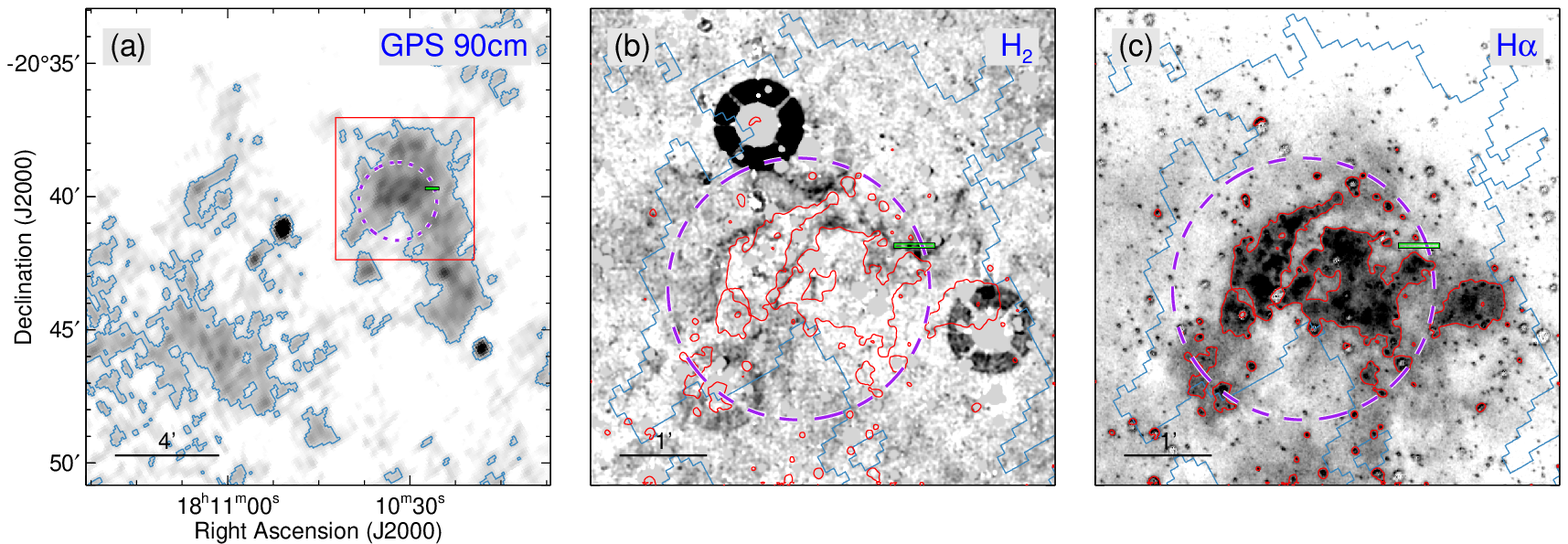}
\caption{
	Multiwavelength images of G9.9$-$0.8:
	(a) radio continuum taken from the VLA 90 cm Galactic Plane Survey
	\citep[GPS 90 cm;][]{hel06},
	(b) continuum-subtracted \hh\ 2.122~\micron\ from the UWISH2 survey
	\citep{fro11,lee19}, and
	(c) continuum-subtracted \ha\ images taken from the SuperCOSMOS \ha\ Survey
	\citep[SHS;][]{par05,stu11}.
	The gray scales of the three images are linear.
	The red box in panel (a) marks the field of view of panels (b) and (c). 
	The blue contours in panels (a)--(c) and
	the red contours in panels (b)--(c) represent
	the 90 cm continuum and \ha\ emission features, respectively.
	The location of the \hii\ region G9.982$-$0.752
	identified by \citet{loc89} is marked by
	a purple circle,
	where the size of the circle corresponds to
	the half-power beam width of his radio observation ($\sim3\arcmin$).
	The green horizontal bar represents the slit position
	of our NIR spectroscopy.
	The bar is made longer than the true slit length ($15\arcsec$)
	for display purposes.
} \label{fig-g99_image}
\end{figure}

\clearpage
\begin{figure}
\includegraphics[width=\textwidth]{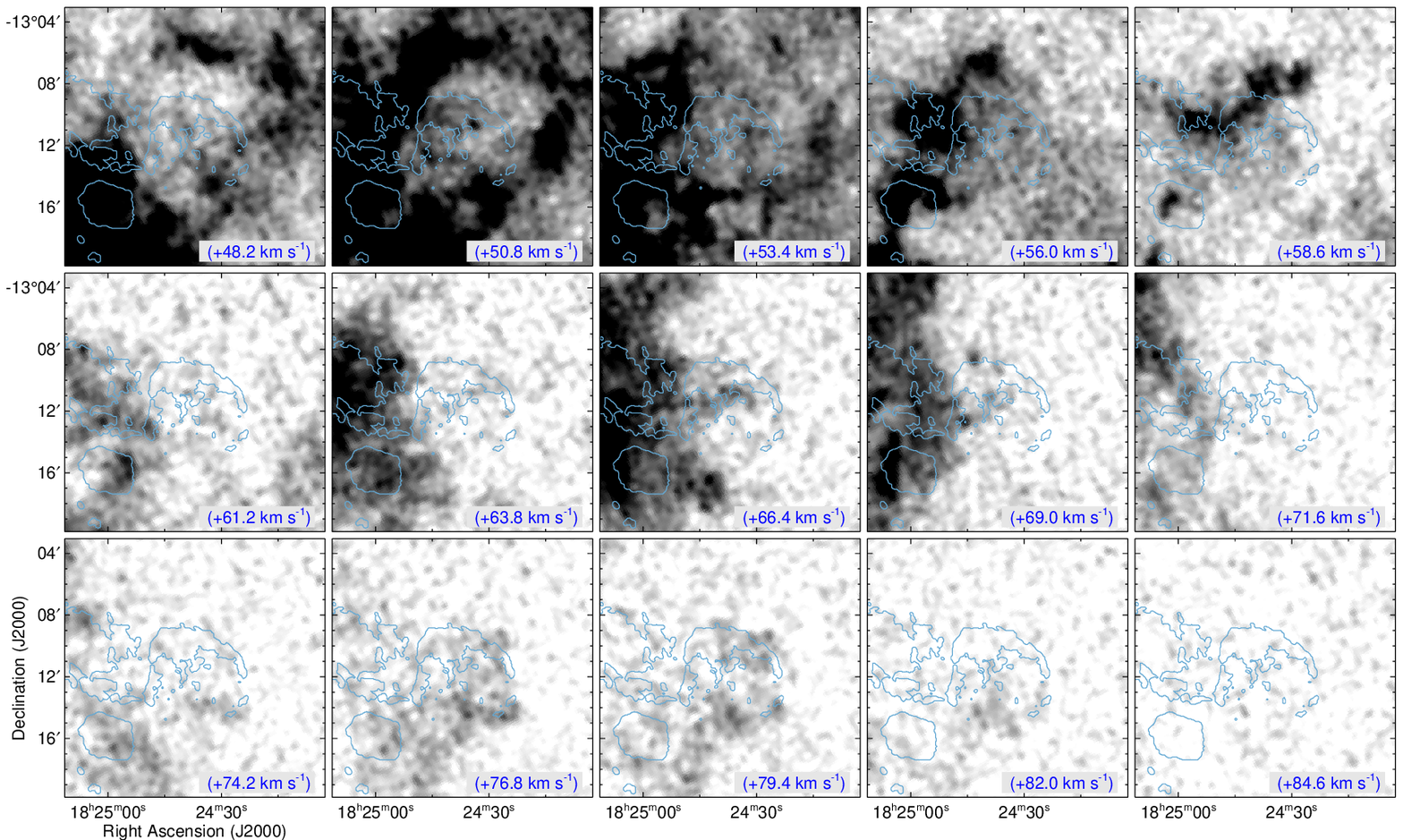}
\caption{
	$^{12}$CO $J=1$--0 channel maps of G18.1$-$0.1 
	in the velocity range of $+48$--$+85~\kms$.
    Each channel map has been obtained by integrating over 2.6~\kms. 
	The color scale is linear with the low and high thresholds of
	0 and 90~K~\kms, respectively.
    The blue contour represents the boundary of the remnant
    in 20~cm continuum (see Figure~\ref{fig-slitpos-1}).
} \label{fig-g181}
\end{figure}

\clearpage
\begin{figure}
\includegraphics[width=0.75\textwidth]{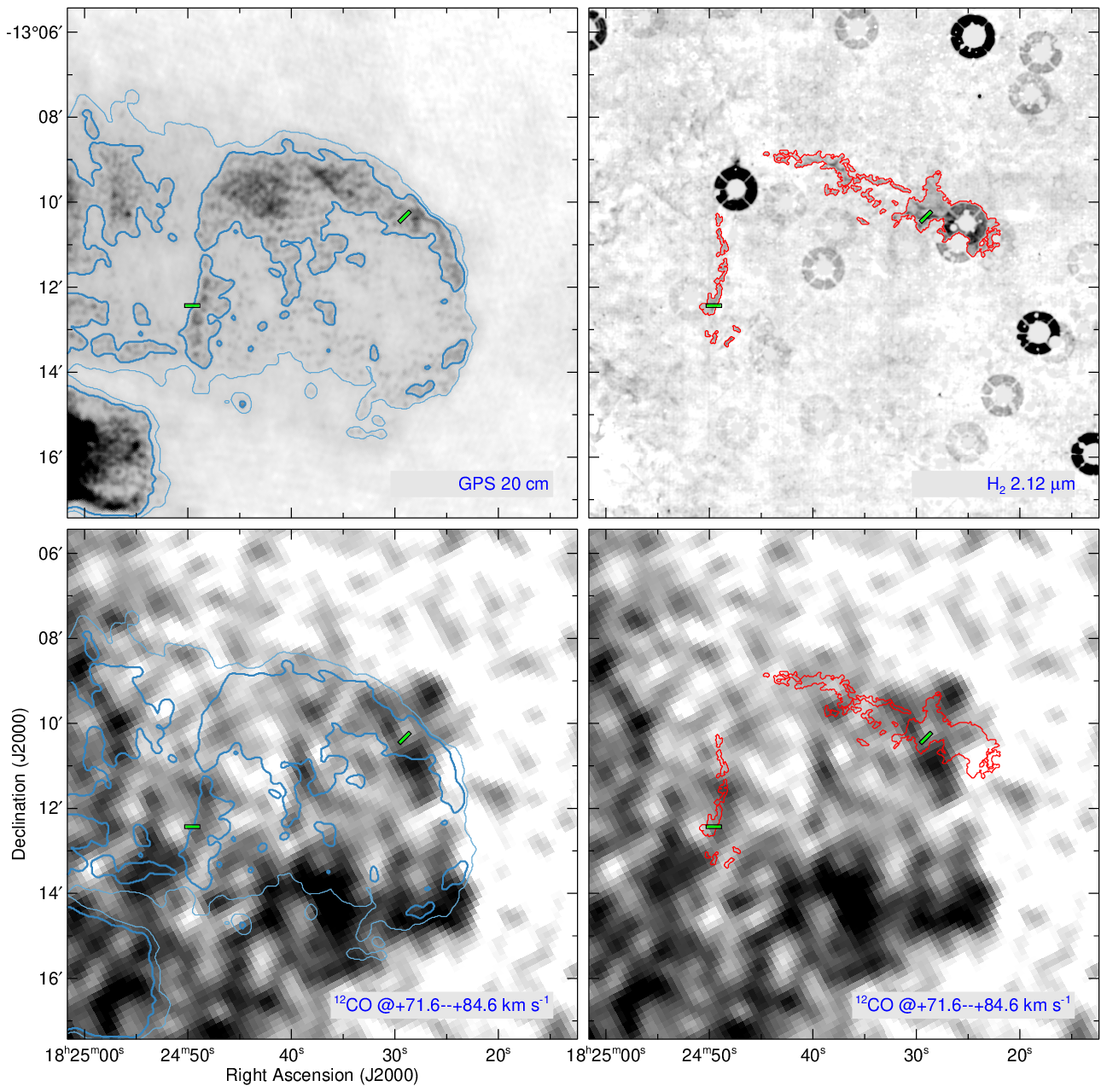}
\caption{
    Top: 20~cm continuum and \hh\ 2.122~\micron\ narrowband images of
    G18.1$-$0.1.
    The blue and red contours show the appearance of the SNR
	in 20~cm continuum and  \hh\ emission, respectively.
	The contour levels in the radio continuum image are
    1.8 and 2.6~mJy/beam.
    The green bars represent the slit positions
	of our NIR spectroscopy.
    Bottom: zoomed-in $^{12}$CO $J=1$--0 intensity maps of G18.1$-$0.1
	integrated from $+72$ to $+85~\kms$.
	The color scale is linear with the low and high thresholds of
	0 and 120~K~\kms, respectively.
} \label{fig-g181_zoom}
\end{figure}

\clearpage
\begin{figure}
\includegraphics[width=0.75\textwidth]{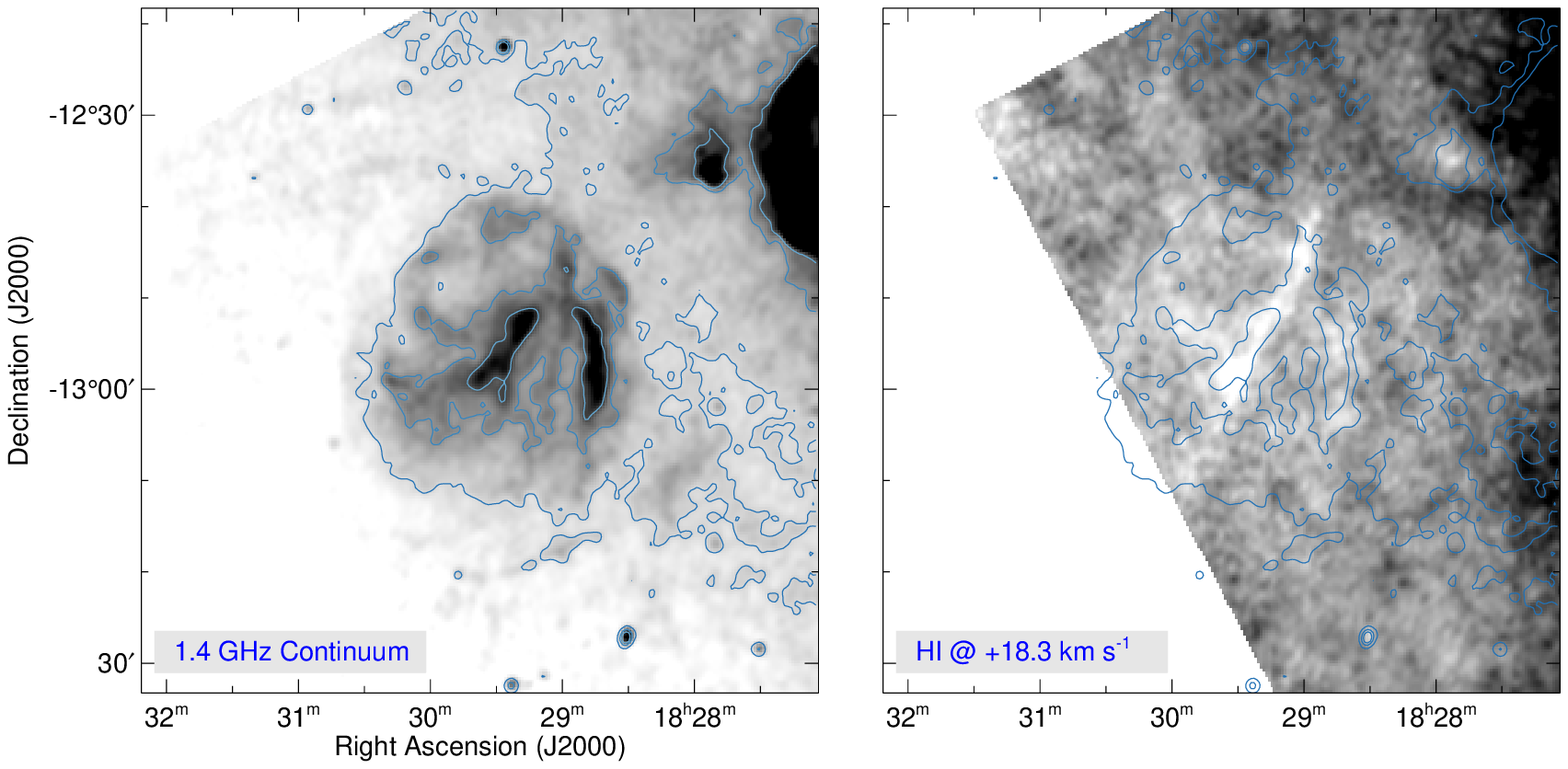}
\caption{
	Left: 21~cm continuum image of G18.9$-$1.1.
	The color scale is linear with the low and high thresholds of
	12 and 30~K, respectively.
	Right: \hi\ 21~cm line channel map at $v_{\rm LSR}=+18.3~\kms$.
	The color scale is linear with the low and high thresholds of
	50 and 110~K, respectively. The blue contours show the appearance of 
	the SNR in 21~cm continuum, and the contour levels are 15, 20, and 25~K.
} \label{fig-g189}
\end{figure}

\clearpage
\begin{figure}
\includegraphics[width=\textwidth]{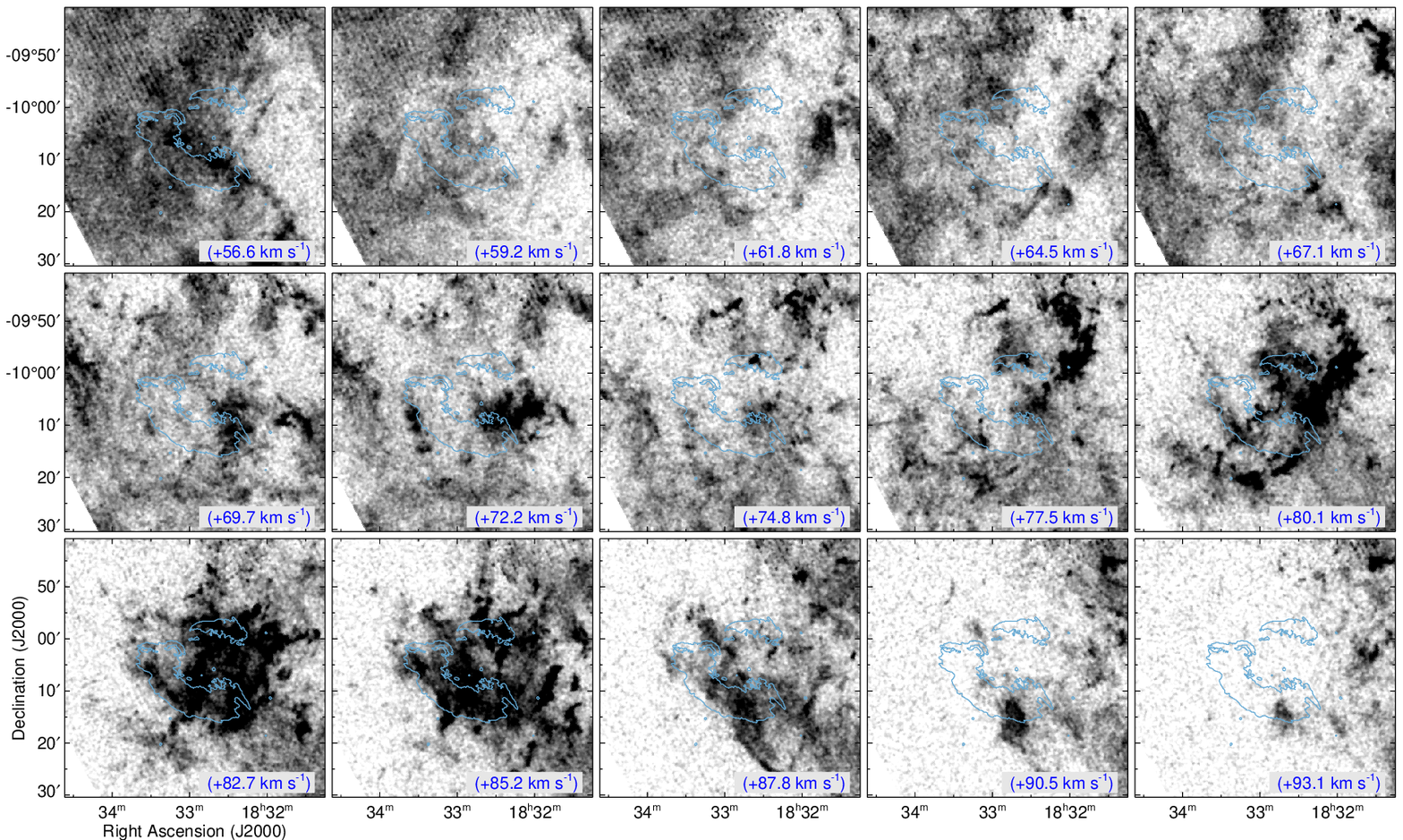}
\caption{   
    $^{12}$CO $J=1$--0 channel maps of Kes 69 
	in the velocity range of $+57$--$+93~\kms$.
    Each channel map has been obtained by integrating over 2.6~\kms. 
	The color scale is linear with the low and high thresholds of
	0 and 60~K~\kms, respectively.
    The blue contour represents the boundary of the remnant
    in 20~cm continuum (see Figure~\ref{fig-slitpos-1}).
} \label{fig-kes69}
\end{figure}

\clearpage
\begin{figure}
\includegraphics[width=0.75\textwidth]{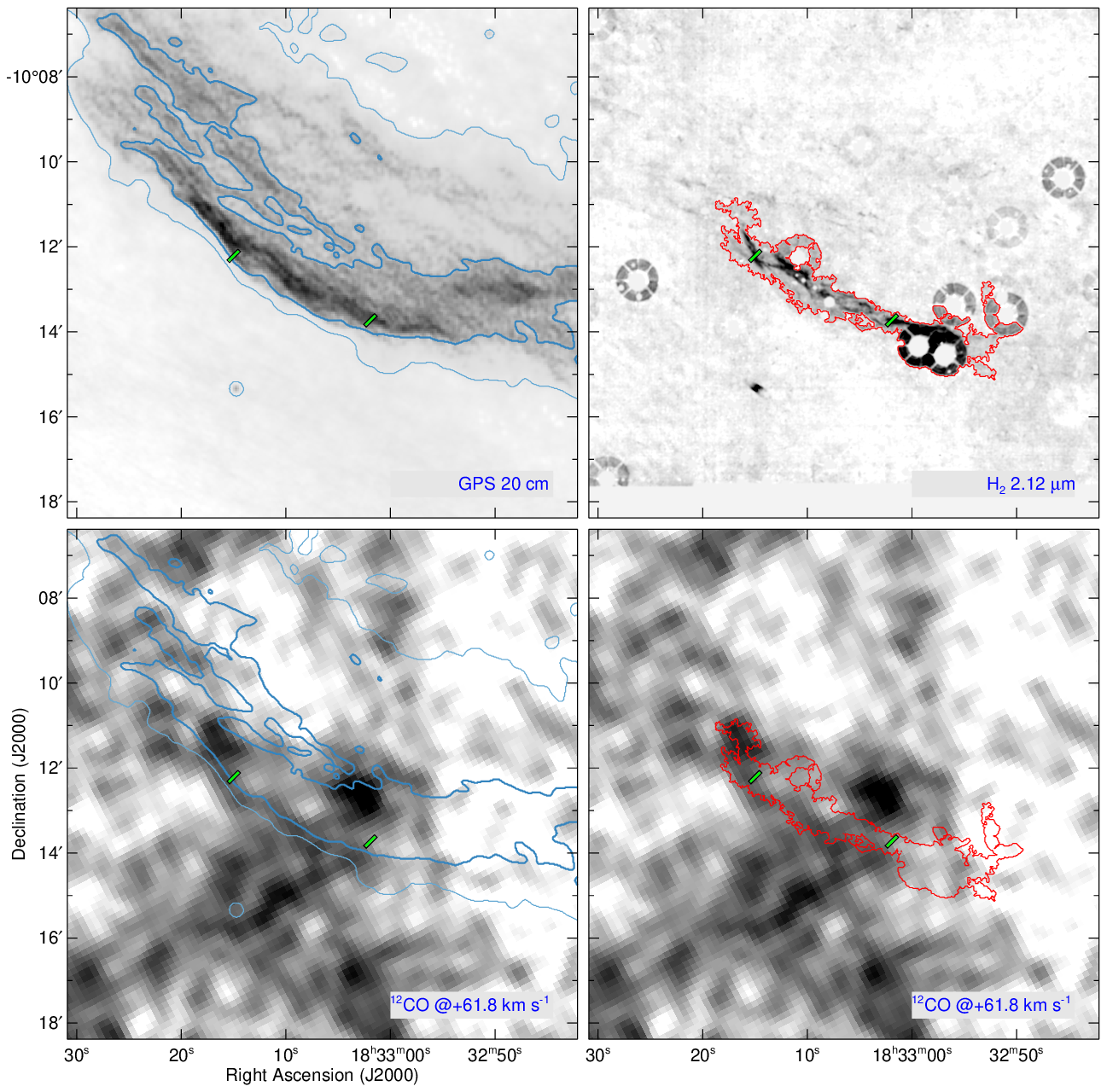}
\caption{
    Top: 20~cm continuum and \hh\ 2.122~\micron\ narrowband images of
    Kes 69.
    The blue and red contours show the appearance of the SNR
    in 20~cm continuum and  \hh\ emission of the SNR, respectively.
    The contour levels in the radio continuum image are
    2 and 6~mJy/beam.
    The green bars represent the slit positions
	of our NIR spectroscopy.
    Bottom: zoomed-in $^{12}$CO $J=1$--0 channel map of Kes 69
    at $+62~\kms$.  
	The color scales are linear, and the low and high thresholds are
	10 and 50~K~\kms, respectively.
} \label{fig-kes69_zoom}
\end{figure}

\clearpage
\begin{figure}
\includegraphics[width=\textwidth]{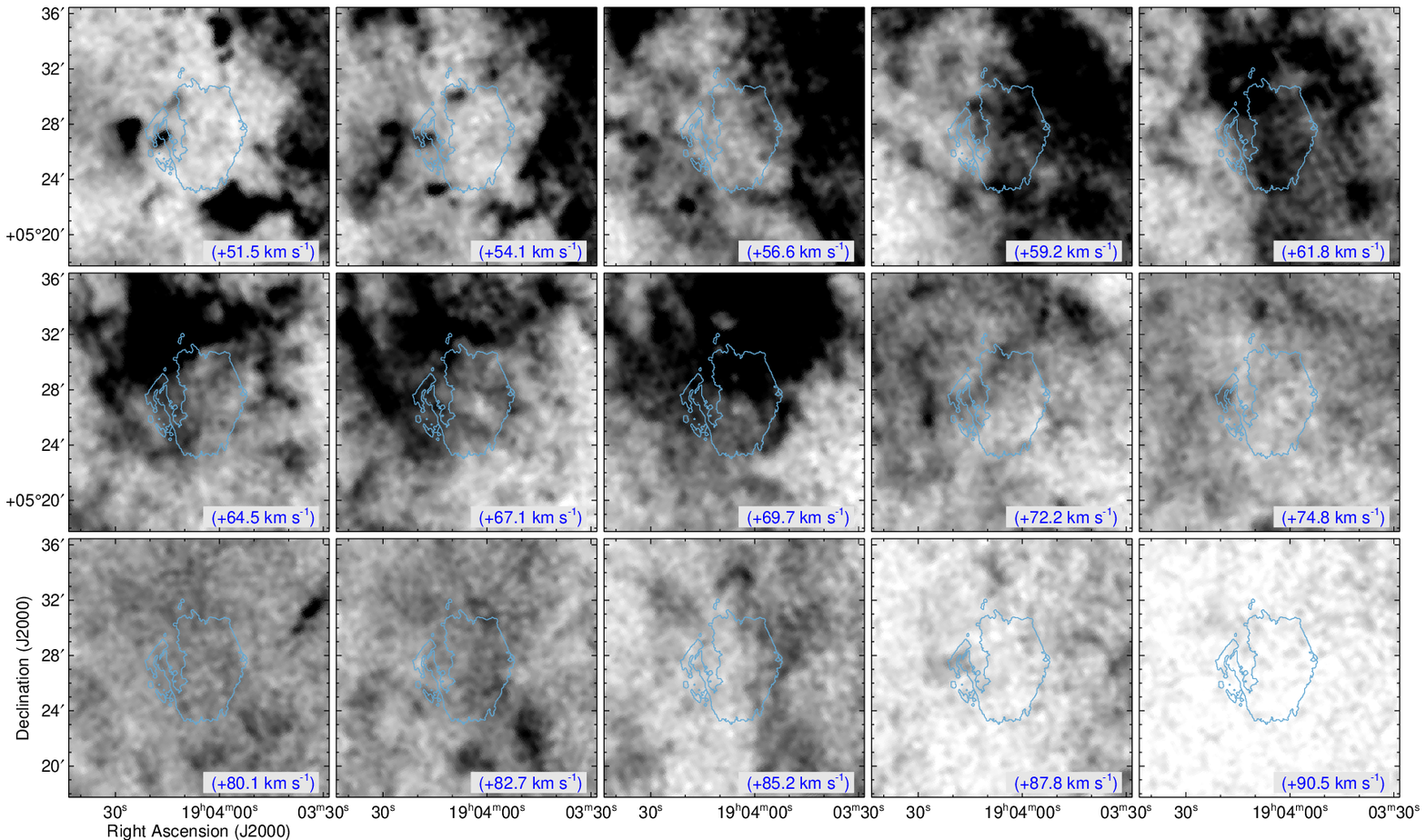}
\caption{
    $^{12}$CO $J=1$--0 channel maps of 3C 396
    in a velocity range of $+52$--$+91~\kms$. 
    Each channel map has been obtained by integrating over 2.6~\kms. 
	The color scales are linear, and the low and high thresholds are
	0 and 60~K~\kms, respectively.
    The blue contour shows the appearance of the SNR 
    in 20~cm continuum (see Figure~\ref{fig-slitpos-1}).
} \label{fig-3c396}
\end{figure}

\clearpage
\begin{figure}
\includegraphics[width=0.75\textwidth]{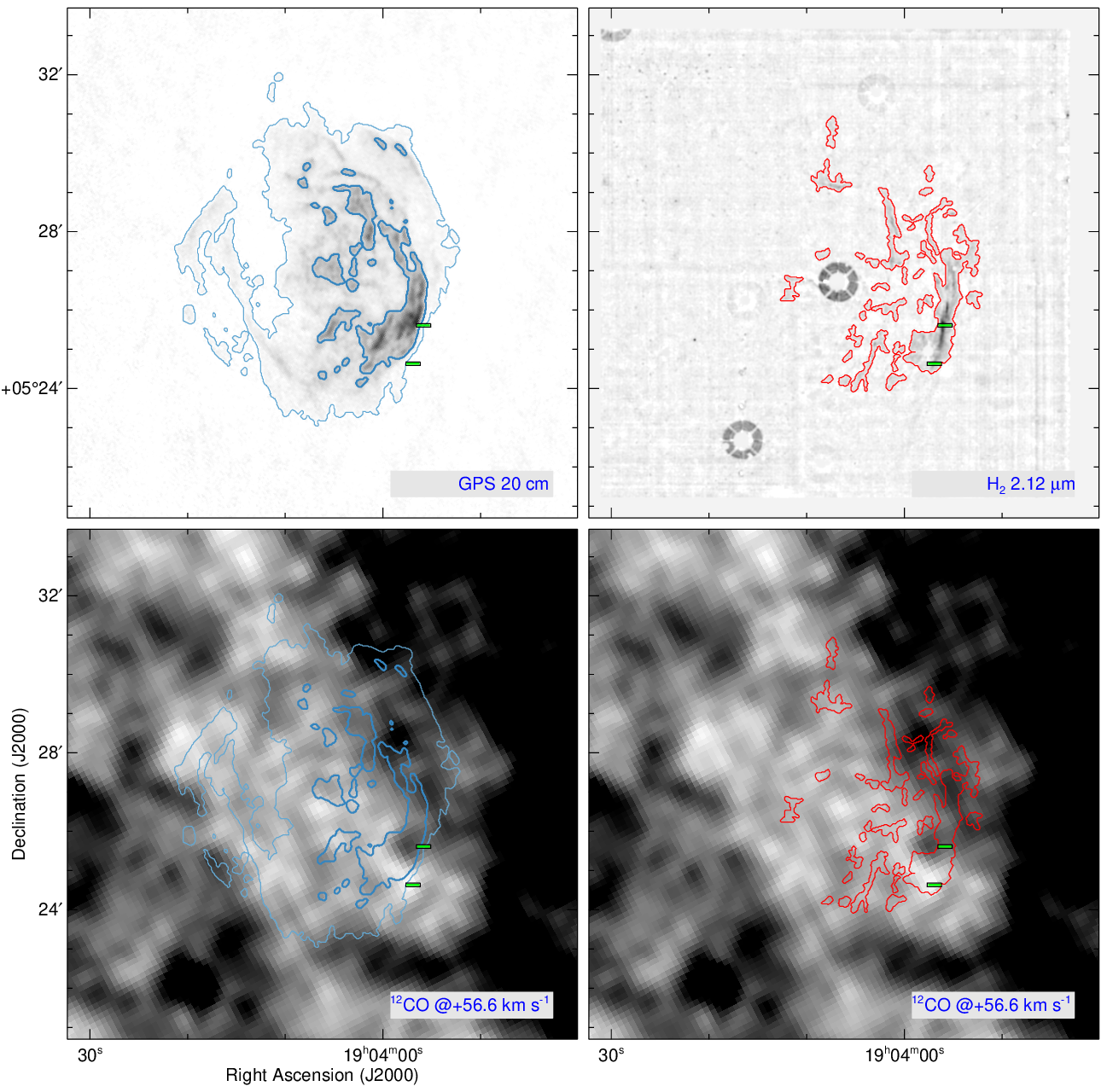}
\caption{
    Top: 20~cm continuum and \hh\ 2.122~\micron\ narrowband images of
    3C 396.
    The blue and red contours show the appearance of the SNR
    in 20~cm continuum and \hh\ emission of the SNR, respectively.
    The contour levels in the radio continuum image are
    0.3 and 3.0~mJy/beam.
    The green bars represent the slit positions
	of our NIR spectroscopy.
    Bottom: zoomed-in $^{12}$CO $J=1$--0 channel map of 3C 396
    at $+56.6~\kms$.  
	The color scales are linear, and the low and high thresholds are
	10 and 50~K~\kms, respectively.
} \label{fig-3c396_zoom}
\end{figure}

\clearpage
\begin{figure}
\includegraphics[width=\textwidth]{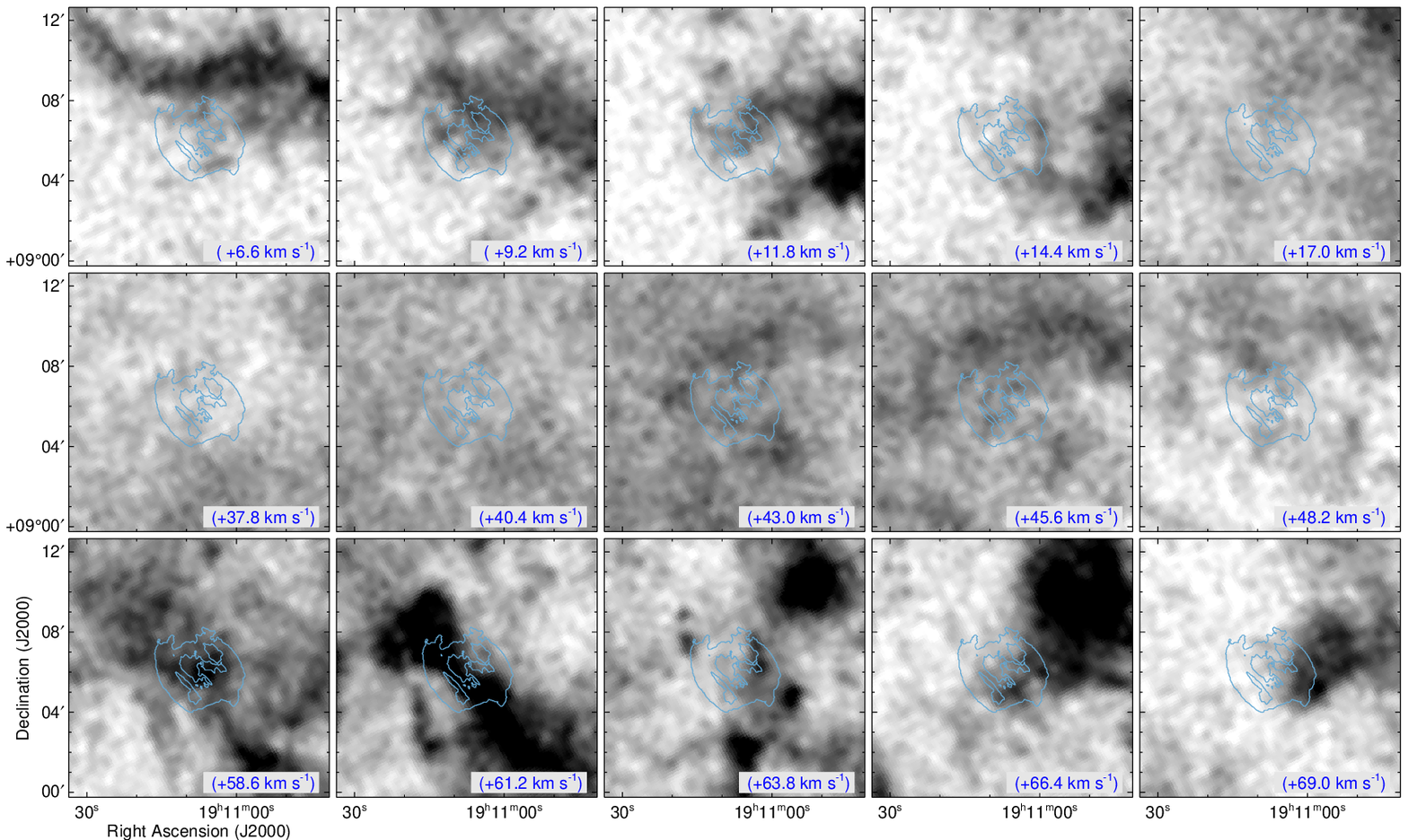}
\caption{
    $^{12}$CO $J=1$--0 channel maps of W49B
    in three different velocity ranges:
    $+7$--$+17~\kms$ (top), $+38$--$+48~\kms$ (middle), and
    $+59$--$+69~\kms$ (bottom).
    Each channel map has been obtained by integrating over 2.6~\kms. 
	The color scales are linear, and the low and high thresholds are
	0 and 80~K~\kms, respectively.
    The blue contour shows the appearance of the SNR 
    in 20~cm continuum (see Figure~\ref{fig-slitpos-1}).
} \label{fig-w49b}
\end{figure}

\clearpage
\begin{figure}
\includegraphics[width=0.75\textwidth]{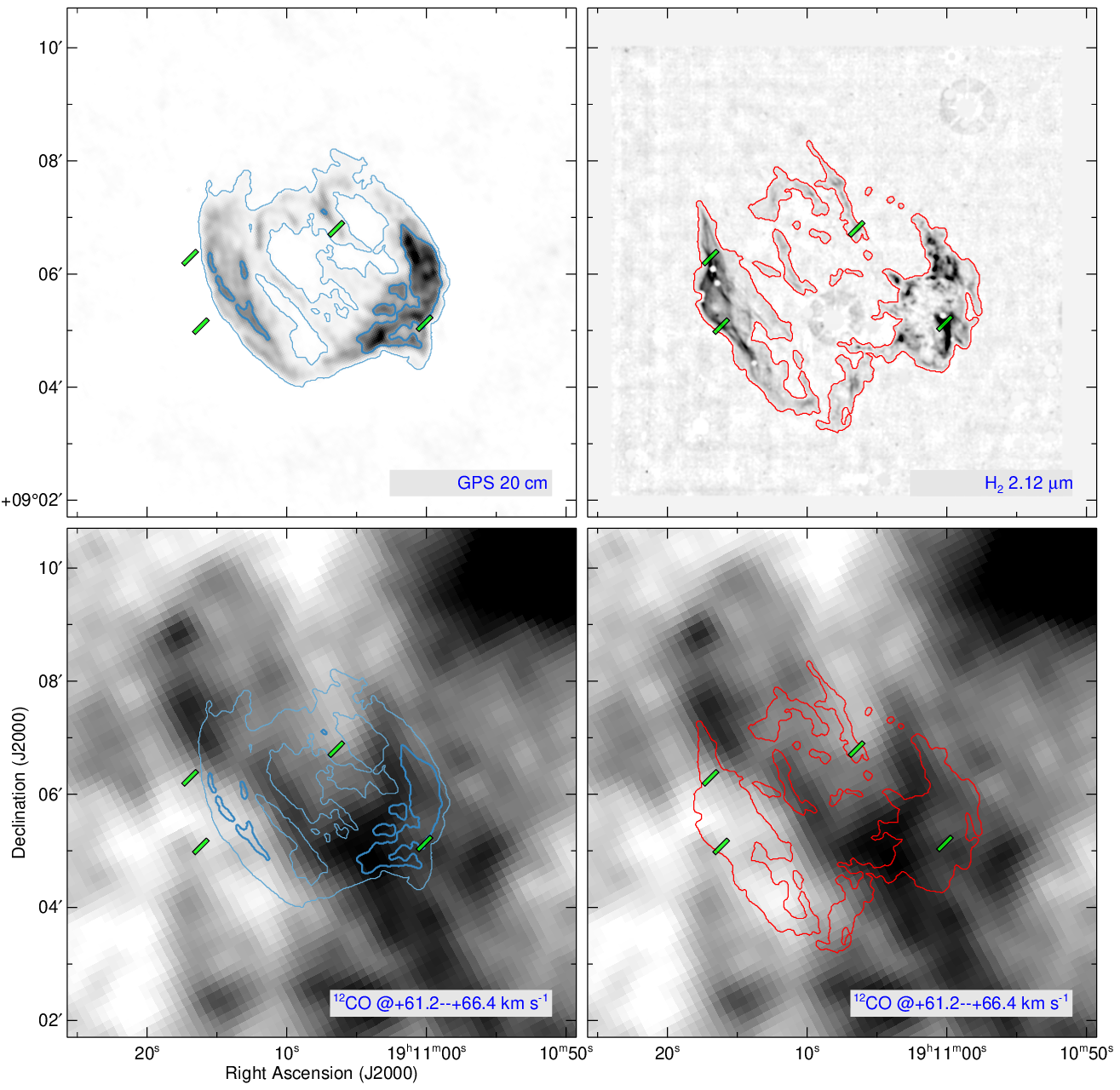}
\caption{
    Top: 20~cm continuum and \hh\ 2.122~\micron\ narrowband images of W49B.
    The blue and red contours show the appearance of the SNR
    in 20~cm continuum and \hh\ emission of the SNR, respectively.
    The contour levels in the radio continuum image are
    3 and 30~mJy/beam.
    The green bars represent the slit positions
	of our NIR spectroscopy.
    Bottom: zoomed-in $^{12}$CO $J=1$--0 intensity map of W49B
    integrated from $+61$ to $+66~\kms$.
	The color scales are linear, and the low and high thresholds are
	40 and 200~K~$\kms$, respectively.
} \label{fig-w49b_zoom}
\end{figure}

\end{document}